\documentclass[a4paper,fleqn,usenatbib]{mnras}
\usepackage[T1]{fontenc}
\usepackage{ae,aecompl}
\usepackage{graphicx}
\usepackage{amsmath}
\usepackage{amssymb}
\usepackage{natbib}
\usepackage{relsize}
\usepackage{enumitem}
\usepackage[utf8]{inputenc}
\usepackage{epstopdf}
\usepackage{subfigure}
\usepackage{pifont}

\usepackage{lipsum}
\usepackage{longtable}

\title[The LDB of NGC~2232]{The {\it Gaia}-ESO survey: A lithium depletion boundary age for NGC~2232}

\author[A. S. Binks et al.]{
A. S. Binks$^{1}$\thanks{E-mail: a.s.binks1@keele.ac.uk},
R. D. Jeffries$^{1}$,
R. J. Jackson$^{1}$,
E. Franciosini$^{2}$,
G. G. Sacco$^{2}$,
A. Bayo$^{3,4}$,
\newauthor
L. Magrini$^{2}$,
S. Randich$^{2}$,
J. Arancibia$^{3,4}$
M. Bergemann$^{5}$,
A. Bragaglia$^{6}$
G. Gilmore$^{7}$,
\newauthor
A. Gonneau$^{7}$,
A. Hourihane$^{7}$,
P. Jofr{\'e}$^{8}$,
A. J. Korn$^{9}$,
L. Morbidelli$^{2}$,
L. Prisinzano$^{10}$,
\newauthor
C. C. Worley$^{7}$ and
S. Zaggia$^{11}$
\\
$^{1}$Astrophysics Group, School of Chemical and Physical Sciences, Keele University, Keele, ST5 5BG, United Kingdom
\\
$^{2}$INAF-Osservatorio Astrofisico di Arcetri, Largo E. Fermi, 5, 50125 Firenze, Italy
\\
$^{3}$Instituto  de  F\'isica  y  Astronom\'ia,  Facultad  de  Ciencias,  Universidad de Valpara\'iso, Av. Gran Breta\~na 1111, Valpara\'iso, Chile
\\
$^{4}$N\'ucleo Milenio de Formaci\'on Planetaria - NPF, Universidad de Valpara\'iso, Av. Gran Breta\~na 1111, Valpara\'iso, Chile
\\
$^{5}$Max-Planck Institut für Astronomie, Königstuhl 17, 69117 Heidelberg, Germany
\\
$^{6}$INAF-Osservatorio di Astrofisica e Scienza dello Spazio via Gobetti 93/3, 40129 Bologna, Italy
\\
$^{7}$Institute of Astronomy, University of Cambridge, Madingley Road, Cambridge CB3 0HA, United Kingdom
\\
$^{8}$N{\'u}cleo de Astronom{\'i}a, Facultad de Ingenier{\'i}a y Ciencias, Universidad Diego Portales (UDP), Santiago de Chile
\\
$^{9}$Observational Astrophysics, Division of Astronomy and Space Physics, Department of Physics and Astronomy, Uppsala University, \\Box 516, SE-751 20 Uppsala, Sweden
\\
$^{10}$INAF-Osservatorio Astronomico di Palermo, Piazza del Parlamento, 1, 90134 Palermo, Italy
\\
$^{11}$INAF-Osservatorio Astronomico di Padova, vicolo dell'Osservatorio, 5, 35122 Padova, Italy
\\
}

\date{Accepted XXX. Received YYY; in original form ZZZ}
\pubyear{2021}

\begin{document}
\label{firstpage}
\pagerange{\pageref{firstpage}--\pageref{lastpage}}
\maketitle

\begin{abstract}
Astrometry and photometry from {\it Gaia} and spectroscopic data from the {\it Gaia}-ESO Survey (GES) are used to identify the lithium depletion boundary (LDB) in the young cluster NGC~2232. A specialised spectral line analysis procedure was used to recover the signature of undepleted lithium in very low luminosity cluster members.  An age of $38\pm 3$ Myr is inferred by comparing the LDB location in absolute colour-magnitude diagrams (CMDs) with the predictions of standard models. This is more than twice the age derived from fitting isochrones to low-mass stars in the CMD with the same models. Much closer agreement between LDB and CMD ages is obtained from models that incorporate magnetically suppressed convection or flux-blocking by dark, magnetic starspots. The best agreement is found at ages of $45-50$\,Myr for models with high levels of magnetic activity and starspot coverage fractions $>50$ per cent, although a uniformly high spot coverage does not match the CMD well across the full luminosity range considered. 
\end{abstract}


\section{Introduction}\label{S_Intro}

Estimating the ages of stars and star clusters is of great importance in astrophysics, but age is something which cannot directly be measured \citep{2010a_Soderblom}. Star clusters, with their populations of nearly-coeval stars of similar initial compositions but with a broad range of masses, offer the most incisive tests of stellar physics and a route towards establishing the time-dependence of physical processes associated with star formation and stellar evolution, and a means of calibrating secondary age indicators (e.g. rotation and abundance ratios, see e.g. \citealt{2019a_Casali,2020a_Casali}) that can be applied to more general galactic populations.

A fundamental test of stellar models, that can be applied very effectively in star clusters, is that ages derived from multiple methods, that are sensitive to different aspects of stellar physics or that sample different parts of the stellar mass spectrum, should agree. In the realm of young stars, there has been growing disquiet that ages determined from high-mass stars evolving on and away from the main sequence are systematically older (by factors of two in the youngest clusters) than the ages determined from fitting isochrones in colour-magnitude diagrams for low-mass pre main sequence (PMS) stars in the same clusters \citep{2006a_Lyra, 2009b_Naylor, 2013a_Bell,2015a_Herczeg, 2016a_Feiden}. 

Further indications of significant problems in the physics of low-mass PMS stars come from discrepancies between isochronal ages and the amount (and dispersion) of lithium that PMS stars deplete as they contract and their cores become hot enough to ``burn'' lithium \citep{2017a_Jeffries,2018a_Bouvier}.  There are also direct and indirect indications that magnetically active stars, whether they are fast-rotating and young or members of close, tidally-locked binary systems, have larger radii than predicted by the most commonly used stellar models \citep{2009a_Morales, 2013a_Torres, 2014b_Malo, 2015a_Kraus, 2017b_Kraus, 2020a_Rizzuto}. This has led to suggestions that rotation, magnetic fields and high surface coverage of starspots may significantly alter the evolutionary tracks and isochrones in young clusters \citep{2013a_Feiden, 2014a_Jackson, 2015a_Somers, 2017a_Macdonald}. If so, this would lead to an underestimate of young cluster ages by factors of $\sim 2$ and a significant underestimate of stellar masses, particularly at low masses, when models that neglect these effects are adopted \citep[e.g.][]{2016a_Feiden, 2017a_Jeffries, 2021a_Macdonald}. 

Resolving the age discrepancies in young clusters is crucial to correctly infer the properties and evolutionary processes of several strongly accreting stars identified at relatively late stages of their PMS, for example, TW Hydrae and Hen 3-600A (\citealt{2000a_Muzerolle}, Ronco et al. in preparation), dozens of accreting low-mass stars in Upper Scorpius (\citealt{2020a_Manara}) and many more ``classical'' $>10\,$Myr old accretors identified in young stellar groups (\citealt{2001a_Haisch,2004a_Mamajek,2013a_De_Marchi,2017a_Beccari,2020a_Silverberg}).

Choosing which models and age scales to adopt is difficult because most age determination methods still have significant physical uncertainties associated with them; for example, the treatment of convection, the amount of core-overshooting and the lack of a detailed understanding of how rotation and magnetic fields influence the stellar structure. It is not clear that any of the ages discussed above are correct! 

Most weight should be attached to methods with the least model-dependence and the ``lithium depletion boundary'' (LDB) technique is presently the least model-sensitive of those available \citep{2013a_Soderblom}. When low-mass, fully-convective, PMS stars contract, their initial Li content is rapidly consumed once their cores reach $\sim 3 \times 10^6$\,K because of the steep temperature dependence of the $^7$Li(p,$\alpha$)$^4$He reaction. More massive PMS stars reach this point more rapidly and the efficiency of convective mixing ensures that this is reflected in their photospheric Li abundance shortly afterwards \citep[e.g.,][]{1997a_Bildsten}. The net effect is that in a cluster of stars with a range of masses, there is predicted to be a sharp transition, the LDB, between stars that have depleted all of their Li and those with only slightly lower masses and luminosities that still retain all their initial Li. The sharpness of this transition persists even after accounting for the known Li dispersion for a given mass/$T_{\rm eff}$ bin.

The luminosity at the LDB therefore has the potential to be a precise age indicator but it is also likely to be accurate. Theoretical parameter studies have varied physical inputs over the range of their remaining uncertainties and found that LDB ages are unlikely to change by more than about 10 per cent at the youngest ages for which the technique is viable and just a few per cent at older ages \citep{2004a_Burke,2015a_Tognelli}. Even the adoption of models that incorporate magnetic activity and radius inflation, which can increase isochronal ages by a factor of 2, result in systematic LDB age increases of just 10-20 per cent in young groups, as predicted by models accounting for the effects of starspots  \citep{2014b_Jackson,2015a_Somers}, and supported by LDB analyses of young groups, e.g., the $\beta$ Pictoris Moving Group \citep[21-26\,Myr,][]{2016a_Binks}.

LDB ages have been established in only about a dozen young clusters and associations with ages between about 20 and 700 Myr \citep[see][and references therein]{2013a_Jeffries, 2013a_Soderblom, 2018a_Martin}. The method requires an assessment of the Li abundances in very low-mass, low luminosity cluster M-dwarfs and thus large amounts of spectroscopic time on large telescopes. Nevertheless, the results are extremely valuable; obtaining a densely sampled set of LDB age determinations in the age range where it is sensitive can identify deficiencies in stellar models and empirically calibrate evolutionary timescales for contracting PMS stars. 

Current results indicate that LDB ages are usually older than those obtained by isochrone fitting \citep{1998a_Stauffer, 1999a_Stauffer, 2005a_Jeffries, 2013a_Jeffries, 2014a_Binks, 2014a_Malo}. This suggests: (i) that high-mass stellar models need to incorporate modest levels of convective core overshoot and/or rotational mixing to provide matched main sequence turn-off ages; and (ii) that standard low-mass isochronal ages may need revising upwards by incorporating new physics into the PMS modelling. 

In this paper we report a new LDB age determination for the young cluster NGC 2232, using spectroscopy obtained as part of the Gaia-ESO spectroscopic survey \citep[][hereafter GES]{2012a_Gilmore,2013a_Randich}. With an age of $\sim 30$ Myr, NGC~2232 is at an interesting stage in its evolution where ages can be estimated from isochronal fits to both low- and high-mass stars as well as the LDB. The technique for selecting NGC~2232 members is described in $\S$\ref{S_Targets}, while $\S$\ref{S_EWLi} explains the method for estimating the relative Li content of the targets. Ages for NGC~2232 are estimated in $\S$\ref{S_LDB} and $\S$\ref{S_CMD}, using the LDB method and fits to low-mass model isochrones respectively. A comparison of these semi-independently derived ages and the implications for early stellar evolution and the cluster age-scale are discussed in $\S$\ref{S_Discussion}.

\section{Target selection}\label{S_Targets}
\subsection{NGC~2232}\label{S_NGC2232}

NGC~2232 is a bright, young open cluster located in  Monoceros. It was first catalogued by \cite{1864a_Herschel} and \cite{1888a_Dreyer} and its distance first estimated by \cite{1931a_Collinder} ($d = 425\,$pc). A spectroscopic study of 16 members by \cite{1974a_Levato} reported $E(B-V) = 0.06 \pm 0.03$\,mag and placed the cluster at $375\,$pc. Photometric studies by \cite{1972a_Claria}~and subsequently by \cite{2006a_Lyra}~reported similar distance ($d = 360$ and $320\pm30$\,pc, respectively), reddening ($E(B-V) = 0.01$ and $0.07\pm0.02$\,mag) and ``nuclear'' (main sequence turn-off) ages ($= 20$ and $32\pm15$\,Myr). Lyra et al. also reported an isochronal age of 25-32\,Myr for low-mass PMS members. 

A chemical abundance analysis of NGC~2232 F and G-type stars by \cite{2010a_Monroe} found a super-solar metallicty of [Fe/H] $= 0.27 \pm 0.08$\,dex and that their (probably still undepleted) Li-abundances were consistent with clusters of $\sim 100$\,Myr or younger. No previous study has focused on Li-depletion in the lower mass stars of NGC~2232. There are 14 NGC~2232 targets (see $\S$\ref{S_TargetSelection}) with spectral-types FGK that were observed in GES and that have high SNR spectra with reported [Fe/H] values and uncertainties (see Appendix~\ref{S_Metallicity}). These metallicities were derived using the same methods as for other young clusters observed in GES \citep[e.g.][]{2017a_Spina} and have been externally verified against Gaia benchmark stars \citep{2014a_Jofre}.  Based on these measurements, Appendix~\ref{S_Metallicity} suggests a near-solar metallicity for NGC~2232 of [Fe/H]$=0.00$ with a dispersion of just 0.05 dex.

\subsection{Selecting NGC~2232 members}\label{S_TargetSelection}

NGC~2232 was observed as part of GES between 6 and 11 November 2015. Targets were selected in CMDs, based on their available optical and near-IR photometry, from a broad region more-than-encompassing the likely location of cluster members. The spectra were recorded with the FLAMES fibre instrument \citep{2002a_Pasquini}, on ESO's UT-2 Very Large Telescope, either with the UVES spectrograph (resolving power, $R\simeq 47\,000$, wavelength range $\lambda\lambda4200-11000$\,\AA) for the minority of bright targets, or with the GIRAFFE intermediate resolution spectrograph ($R\sim 12\,000, \lambda\lambda 3700-9000$\,\AA). Both setups include the Li~{\sc i}~6708\AA\ absorption line. Raw images were homogeneously analysed and spectra extracted and calibrated using standard GES pipelines \citep[see][]{2014b_Jeffries, 2014a_Sacco,2018a_Randich}. The spectra in this paper are from the fifth internal data release (GESiDR5\footnote{The GESiDR5 catalogue is available to the GES consortium at http://ges.roe.ac.uk/.}).

A list of high probability members of NGC~2232 with GES spectroscopy was taken from \cite{2020a_Jackson}. This study assigns membership using astrometry from Gaia DR2 (\citealt{2018a_Gaia_Collaboration}) and radial velocity (RV) measurements from GESiDR5. Since the membership probabilities are calculated using kinematics alone, the target list is unbiased with respect to the presence of lithium or indeed any age-related property. Of the 760 targets with spectroscopy, 80 have a membership probability $P_{\rm 3D} > 0.95$, with an average value of $0.992$. Based on these probabilities we expect $\leq 1$ contaminant in our list of members.

Since the publication of \cite{2020a_Jackson}, the (early) Third Gaia Data Release has been made available (herein Gaia EDR3, \citealt{2020a_Gaia_Collaboration}). The statistical uncertainties for parallax measurements in Gaia EDR3 are typically $\sim 30$ per cent smaller and, more importantly, the systematic uncertainties due to possible correlated errors in the parallax zero-point on small spatial scales (\citealt{2018a_Lindegren}) have been significantly improved (\citealt{2020a_Lindegren}). Although the membership probabilities use Gaia DR2 astrometry, we measure the weighted mean cluster parallax ($\pi_{\rm c}$) and distance modulus ($d_{\rm mod}$) using Gaia EDR3 parallaxes as follows: first an intrinsic dispersion of the cluster parallax is estimated, equal to the standard deviation of cluster members minus the RMS parallax uncertainty (subtracted in quadrature). This dispersion is added in quadrature with the parallax uncertainties from each target and used as a weight to give a mean $\pi_{\rm c} = 3.1355\pm0.0106\pm0.0300$\,mas and $d_{\rm mod}=7.518\pm0.007\pm0.021$\,mag, where the two error bars represent the statistical error in the mean and a remaining systematic uncertainty (see figure 2a in \citealt{2020a_Lindegren}), respectively. The corresponding distance of $d=319\pm1\pm3$\,pc, is very similar to \cite{2006a_Lyra} but with a much smaller error bar. We discuss how the distance measurement affects our analyses in $\S$\ref{S_LDB} and $\S$\ref{S_CMD}.

We use Gaia DR2 $G$-band (optical) and $K_{\rm s}$-band (near-IR) photometry for our analyses. This is because Gaia provides homogeneous $G$ magnitudes with mmag-precision for all targets in our sample (our faintest target has $G\sim19$) and absolute $K_{\rm s}$ magnitudes are preferable in identifying the LDB since they are highly sensitive to the peak flux from low-mass stars. Near-IR photometry is from 2MASS (\citealt{2003a_Cutri}) and the sixth data release of the Vista Hemisphere Survey (VHS, \citealt{2019a_McMahon}). There are 2MASS measurements available for all stars, and all these have the best possible flags for quality, contamination and confusion (in all three bands). $K_{\rm s}$ magnitudes are also available for every target in VHS but bright targets have saturated at $K_{\rm s}$. For 49 sources where the 2MASS $K_{\rm s}<13$ then that value is used. For 31 fainter sources, the VHS $K_{\rm s}$ photometry is adopted.

The $G-K_{\rm s}$ colours are dereddened using the $E(B-V)$ value calculated in \cite{2006a_Lyra}, $R_{V}=3.09$ and $A_{K_{\rm s}}/A_{V}=0.114$ \citep{1989a_Cardelli}, where $A_{G}$ is estimated using the following fit to the $A_{G}$ versus $G-K_{\rm s}$ relation for main sequence stars provided in \citet[][see their figure 5, top-middle panel] {2018a_Danielski}:

\begin{equation}
A_{G} = A_V\times\frac{0.84-0.04(G-K_{\rm s})+A_{K_{\rm s}}}{(1.0-0.04A_{V})}
\end{equation}

\section{Lithium equivalent widths}\label{S_EWLi}

\subsection{GESiDR5}\label{S_EWLi_GESiDR5}

The equivalent width of the Li~{\sc i}~6708\AA\ feature, EW(Li), is reported as part of the standard GESiDR5 analysis for all but 2 targets in our list. The method of measurement is described in \cite{2016a_Bouvier} and \cite{2018a_Randich}.

\begin{figure}
    \centering
    \includegraphics[width=.48\textwidth]{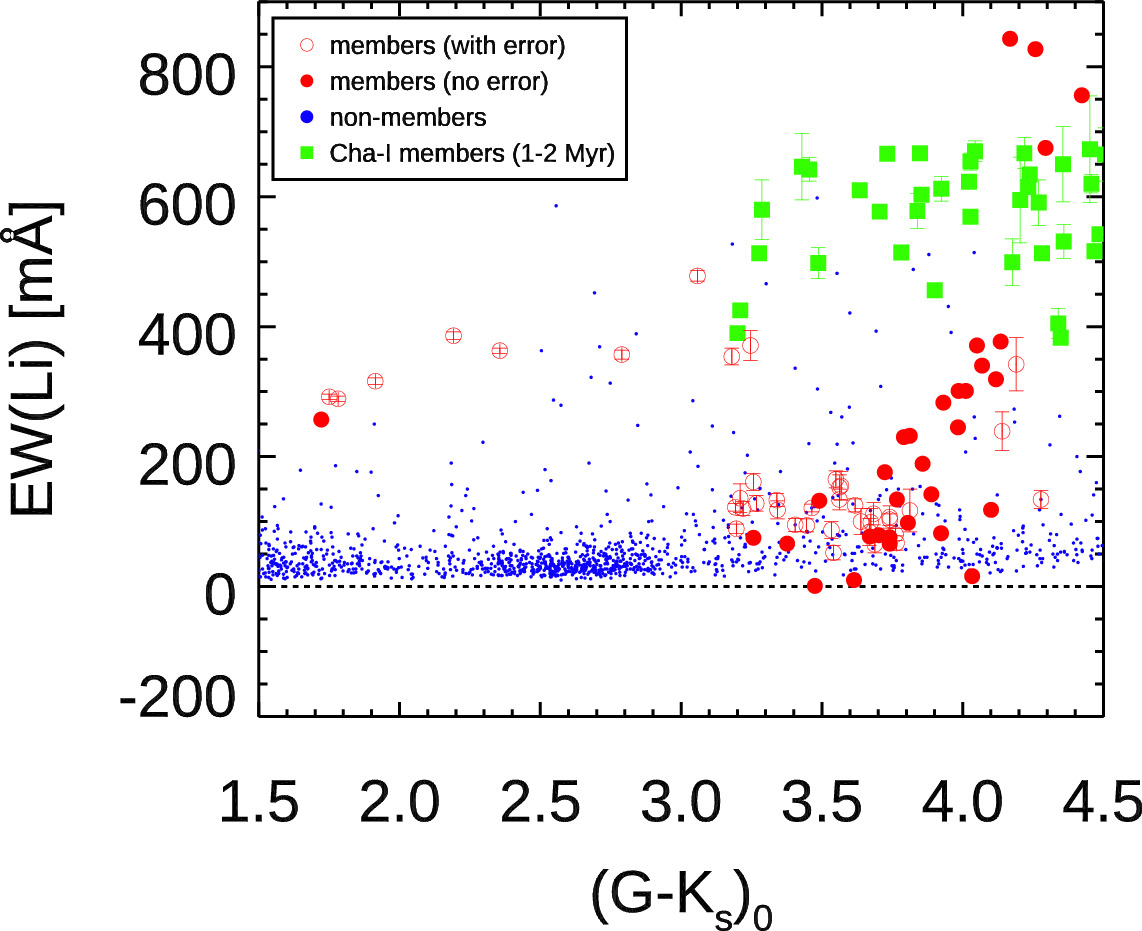}
    \caption{EW(Li) versus $(G-K_{\rm s})_0$ in NGC~2232, using EW(Li) measurements from GESiDR5. Red symbols represent the 80 high probability members ($P_{\rm 3D}>0.95$). Open and filled symbols denote EW(Li) values with and without reported uncertainties in GESiDR5, respectively. Blue points are likely {\it non}-members of NGC~2232 ($P_{\rm 3D} < 0.1$) observed in the same instrument configurations as the members. The green squares are members of the $\sim 2\,$Myr Cha I association also observed as part of GES.}
    \label{fig:ewli_gks}
\end{figure}

The morphology of Figure~\ref{fig:ewli_gks} suggests we are seeing the transition from Li-depleted stars at $(G-K_{\rm s})_0 <4.0$ to Li-rich stars at $(G-K_{\rm s})_0 >4.1$ that marks the LDB. The Li-rich stars have EW(Li) values comparable to those seen in $P_{\rm 3D}>0.9$ members of the Cha~I association (also taken from GESiDR5). Since Cha~I has an age of only $\sim 2$ Myr \citep{2007a_Luhman} then this level of Li likely represents the undepleted local cosmic value.

Whilst these initial indications are promising, there are features in the plot that are at odds with expectations for the EW(Li)/colour distribution of a young open cluster. Firstly, stars immediately hotter than the LDB should be almost completely depleted of Li and their EW(Li) values should be $\sim 0$. However, there appears to be a plateau, with mean EW(Li)$\simeq 120$m\AA, that is also present. This may be indicative of a systematic offset resulting from the EW estimation process in GESiDR5, which is supported by the fact that the EW(Li) of cluster non-members, which are almost certainly depleted of Li, also have EW(Li)>0. The process of measuring EW(Li) is complicated in M-dwarfs by molecular absorption features \citep[e.g.][]{2014a_Rajpurohit} and the pseudo-continuum is highly sensitive to small temperature changes. Errors in continuum placement can easily lead to systematic shifts in EW(Li), particularly in low signal to noise ratio (SNR) spectra. Secondly, most of the targets with $(G-K_{\rm s})_0>3.7$ do not have a reported error-bar in GESiDR5. Given the SNR for these objects is low ($\lesssim 10$, see Figure~\ref{fig:spec}), it is important to quantify the EW(Li) uncertainties to determine whether the targets with large EW(Li) values are real detections or simply cases of noisy data or uncertain pseudo-continuum placement.

\subsection{Remeasuring EW(Li)}\label{S_EWLi_us}

To address the issues identified with the GESiDR5 measurements, we independently measured EW(Li) using the reduced GESiDR5 spectra and a novel technique which is detailed in Appendix~\ref{S_EWLiMethod}. In brief, template Li-free spectra for a given $(G-K_{\rm s})_0$ colour, were generated over the 6675--6730\AA\ range by making use of the large number of field stars serendipitously observed in the fields of GES clusters (i.e., those with kinematic cluster membership probabilities $P_{\rm 3D}<0.1$). For M-dwarfs we can safely assume that the vast majority of these stars have no lithium at all. This is an entirely empirical approach to determining a continuum for cool stars and not reliant on the fidelity of model atmospheres. The FWHM of any Li feature is determined by the instrumental resolution and the rotational velocity of the star given by the {\tt VROT} parameter in GESiDR5. EW(Li) is then estimated by comparing the target spectrum with the template that best matches its intrinsic colour and integrating over a Gaussian profile that characterises the FWHM. Uncertainties are calculated by repeating this procedure for regions in the vicinity of the Li feature. 

For some stars with low SNR  there are no {\tt VROT} values reported in GESiDR5. In these cases we assumed a value of $32\,{\rm km\,s}^{-1}$, which is the mean value for the cluster M-dwarfs with {\tt VROT}. Whilst the FWHM for these stars may be incorrect, the effects on the estimated EW(Li) are smaller than the error bars due to spectral noise and continuum placement. A visual inspection confirmed that the template continua appear well matched to the observations in all cases.

\begin{figure}
    \centering
    \includegraphics[width=.48\textwidth]{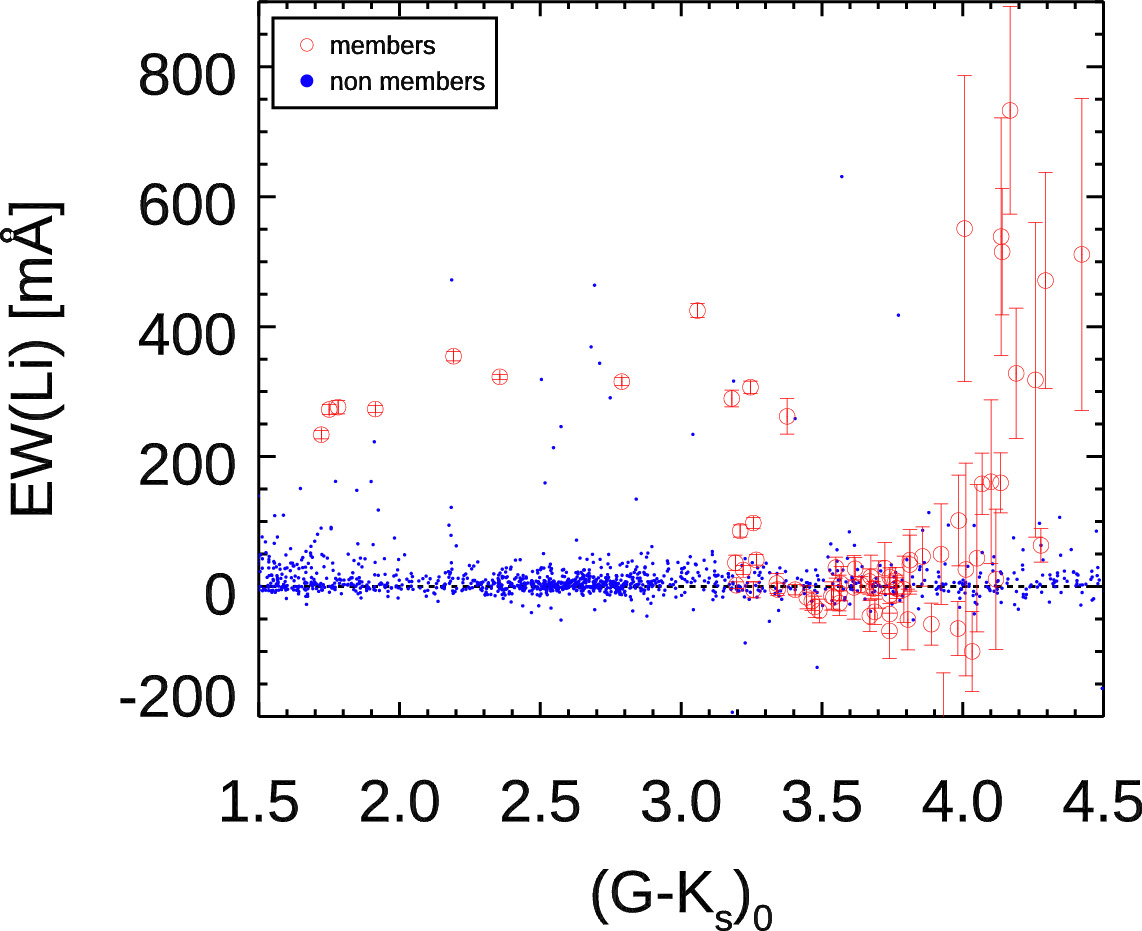}
    \caption{EW(Li) versus $(G-K_{\rm s})_0$, using the revised EW(Li) values (and errors) calculated from our analysis in $\S$\ref{S_EWLi_us}. The symbols and notation are the same as those described in Figure~\ref{fig:ewli_gks}. }
    \label{fig:ewli_gks_rjj}
\end{figure}

\begin{figure*}
    \centering
    \includegraphics[width=.9\linewidth, height=20cm]{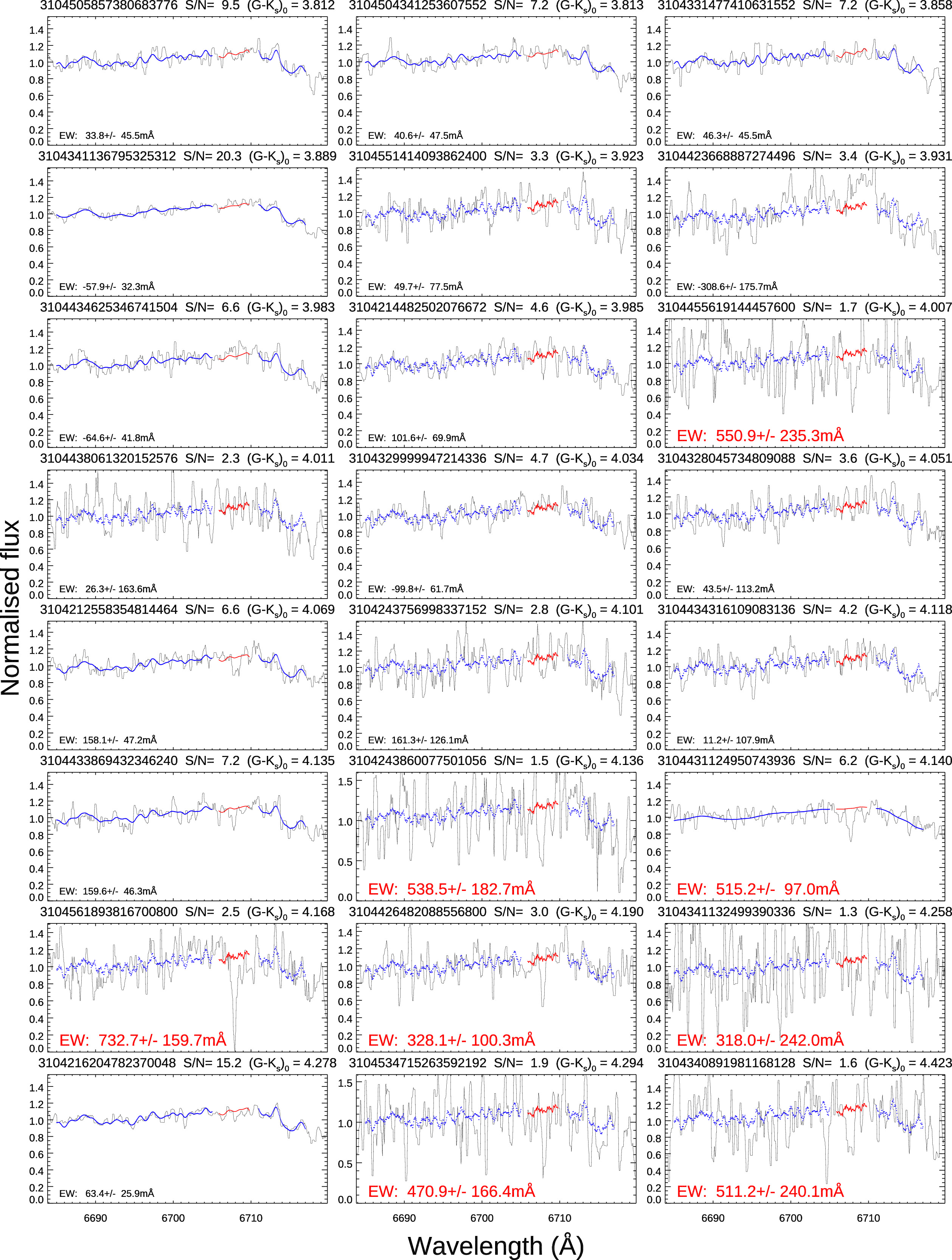}
    \caption{GES spectra for the 24 targets with $(G-K_{\rm s})_0>3.8$. The normalised, reduced spectra from GESiDR5 are displayed in grey, where the data have been binned by 7 pixels. The best-matching spectral template (described in $\S$\ref{S_EWLi_us}) is shown in the region encompassing the Li-feature at 6708\AA\ (in red) and in two regions either side of the feature (in blue). The title of each panel gives the Gaia DR2 source identifier, the SNR given in GESiDR5 and the $(G-K_{\rm s})_0$ colour. The EW(Li) values are provided in the bottom-left of each panel, where targets defined as Li-rich are highlighted in larger red text.}
    \label{fig:spec}
\end{figure*}

The EW(Li) versus $(G-K_{\rm s})_0$ plot for these new EW(Li) estimates is provided in Figure~\ref{fig:ewli_gks_rjj} and in Figure~\ref{fig:spec}~we present the spectra for all NGC~2232 members with $(G-K_{\rm s})_0>3.8$. The issues highlighted in $\S$\ref{S_EWLi_GESiDR5} appear to be resolved with our methods. The median EW(Li) value of targets in the Li-chasm ($3.3<(G-K_{\rm s})_0<4.0$) is $-5\pm72\,$m\AA, as expected for stars that have no Li. Secondly, our analysis provides uncertainties for all targets, regardless of their SNR. 

\section{The LDB of NGC~2232}\label{S_LDB}

This section describes how the LDB is located in NGC~2232 ($\S$\ref{S_LDBident}), the evolutionary models used in this work ($\S$\ref{S_EvoModels}) and the method to calculate an LDB age ($\S$\ref{S_LDB_age}). All data used in this work: $P_{\rm 3D}$, photometry and EW(Li) values from both GESiDR5 ($\S$\ref{S_EWLi_GESiDR5}) and our own analysis ($\S$\ref{S_EWLi_us}) are provided in Table~\ref{T_Data}.

\subsection{Identifying the LDB location}\label{S_LDBident}
Curve of growth models predict that a 99 per cent Li-depleted early/mid M-type star has an EW(Li)$\approx$300\,m\AA~(\citealt{2007a_Palla}), compared  with 600-700\,m\AA\ for no depletion. Therefore we use EW(Li)$=300$\,m\AA\ to discriminate between Li-rich and Li-poor stars. Figures~\ref{fig:CMD_SSM} shows the intrinsic $(G-K_{\rm s})_0/M_K$ CMD for the NGC 2232 members. Symbols are colour coded for whether EW(Li) is bigger or smaller than 300\,m\AA. Objects that have EW(Li) within one error bar of this threshold are shown as open symbols and triangles indicate objects, which by virtue of their position in the CMD, are likely to be unresolved binaries (the exact criterion is discussed in  $\S$\ref{S_CMD}).

The CMD show a reasonably clear boundary between Li-rich and Li-poor targets. There are few targets near the EW(Li) threshold, as expected, since the depletion of Li is rapid once it begins. Looking just at the "single star" sequence we put the LDB somewhere in the grey, rectangular region separating Li-poor from Li-rich stars. The upper bound is defined by the faintest clear Li-poor star, the lower bound is defined by the brightest clear Li-rich star that lies redward of the Li-poor marker. The box width is defined by their separation in colour. Whilst this latter choice is somewhat arbitrary, changing the box {\it width} by factors of two does not significantly affect the calculated ages (see  $\S$\ref{S_LDB_age}). There are two Li-rich stars that are more luminous than the defined LDB box, but these are most likely to be unresolved binaries that are displaced upwards by up to 0.75 mag in the CMD.

\subsection{Evolutionary models}\label{S_EvoModels}
The evolutionary models adopted in this work are categorised as either ``standard models'' or ``magnetic models''. Standard models feature only convective mixing and do not take any account of the influence of magnetic activity on stellar structure. There are numerous standard models that differ in their input physics regarding the equation of state, treatment of convection, interior opacities and atmospheres. As representatives of these, we consider the models of \citet[][herein, D08]{2008a_Dotter}, \citet[][herein, B15]{2015a_Baraffe} and the spot-free models of \citet[][herein, S20]{2020a_Somers}.

Magnetic models incorporate some aspects of the dynamo-generated magnetism which is known to be present in these young, fast-rotating, magnetically active stars (\citealt{2009a_Reiners}). Two evolutionary codes are considered: (a) The S20 models which incorporate the blocking of flux by dark, magnetic starspots at the stellar photosphere. These are available in increments of spot-coverage fractions ($f_{\rm sp}$) of 0.17. The spots are taken to be at a temperature that is 80 per cent of the unspotted photosphere, meaning that about $0.59f_{\rm p}$ of the radiative flux from the star is blocked by the starpots. The spot-free version is used as a standard model. (b) The ``magnetic Dartmouth models'' described in \cite{2014a_Feiden} and \citet[][herein F16]{2016a_Feiden}, which implement magnetic inhibition of convection constrained by a boundary condition of an average $2.5\,$kG magnetic field at the stellar surface, which is approximately the equipartition value at the surface of a mid-M dwarf. These models are approximately an extension of the D08 standard model.

For consistency, the $(G-K_{\rm s})_{0}$ and $M_{\rm K_{\rm s}}$ values from each model are calculated from $\log T_{\rm eff}$ and $\log L$, using the same, age-dependent cubic relationships between colours and temperatures and between bolometric corrections and luminosities, derived from the models of \cite{2015a_Baraffe}. Specific relationships were calculated at each age between 1.0 and 250.0\,Myr (in steps of 0.1\,Myr). This was done to remove any disparities due to the adoption of different bolometric corrections in the native colour and magnitude predictions of each model. For the S20 models there is the added complexity in accounting for two photospheric temperatures: one from the cooler star-spot regions and another from the warmer surrounding photosphere. The $G$ and $K_{\rm s}$ bolometric corrections in this case were calculated using equation 6 in \cite{2014b_Jackson}, where the temperature ratio between the spotted surface and the photosphere was fixed at 0.8.

\subsection{Estimating the LDB age of NGC~2232}\label{S_LDB_age}
\begin{figure}
\centering
  \includegraphics[width=.48\textwidth]{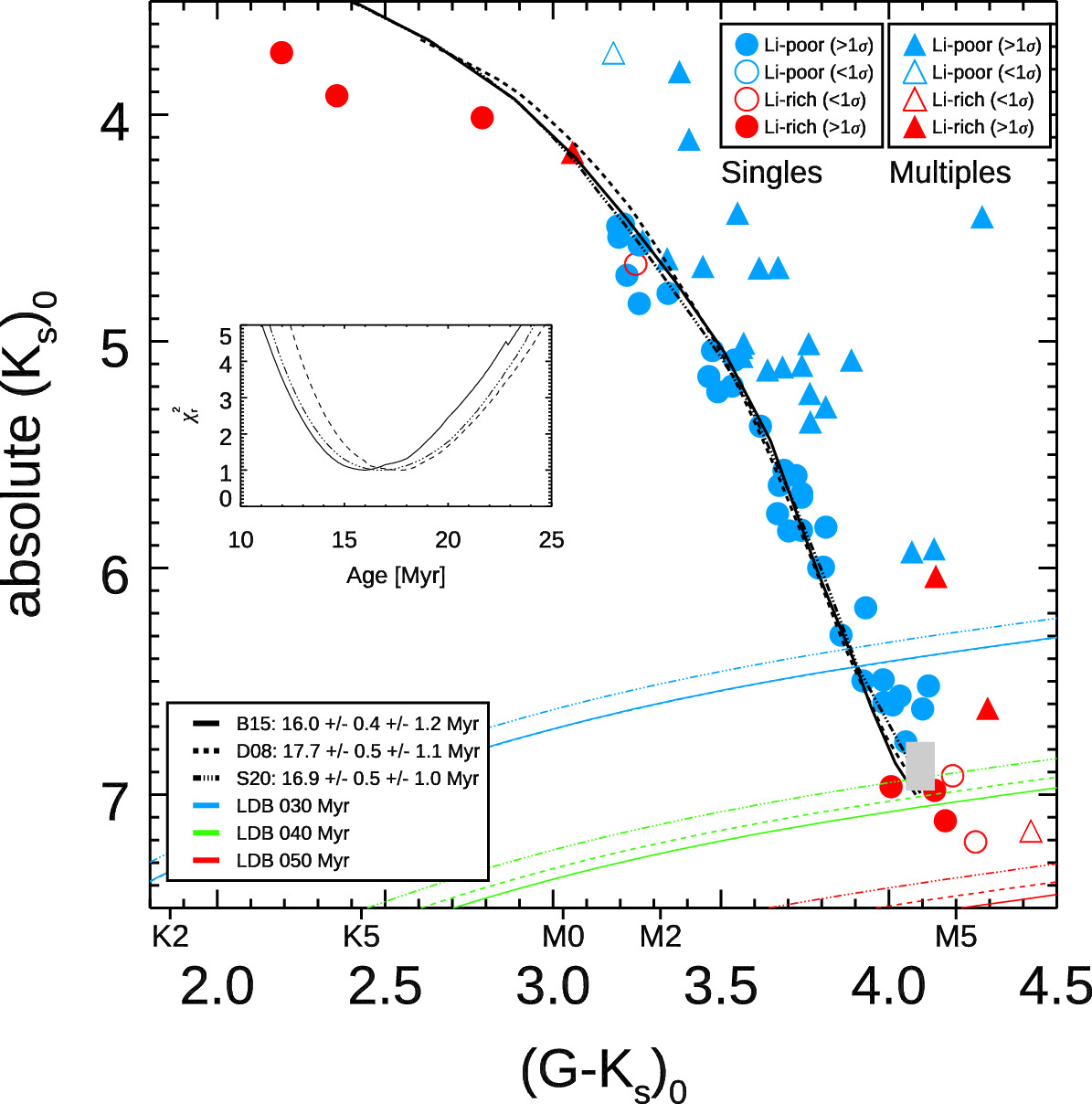}
\caption{$M_{K_{\rm s}}$ vs ($G-K_{\rm s})_0$ for high probability cluster members. Red and blue symbols denote targets classed as Li-rich and Li-poor (in $\S$\ref{S_LDBident}), where open and closed symbols denote whether the EW(Li) is within, or not within, one error bar of 300\,m\AA. Circles and triangles represent the likely-single and likely-multiple stars (in $\S$\ref{S_CMD}), respectively. The grey box denotes the estimated position of the lithium depletion boundary (LDB). The blue, green and red lines represent curves of constant luminosity at 99 per cent Li depletion from three standard models (see $\S$\ref{S_EvoModels} for details) at 30, 40 and 50\,Myr, respectively. The black curves are the best-fitting $M_{K_{\rm s}}$ vs $(G-K_{\rm s})_0$ isochrones to the likely single stars for the same models.
The inset plot shows the reduced $\chi^{2}_r$ as a function of isochrone age.}
\label{fig:CMD_SSM}
\end{figure}

\begin{figure}
\centering
  \includegraphics[width=.48\textwidth]{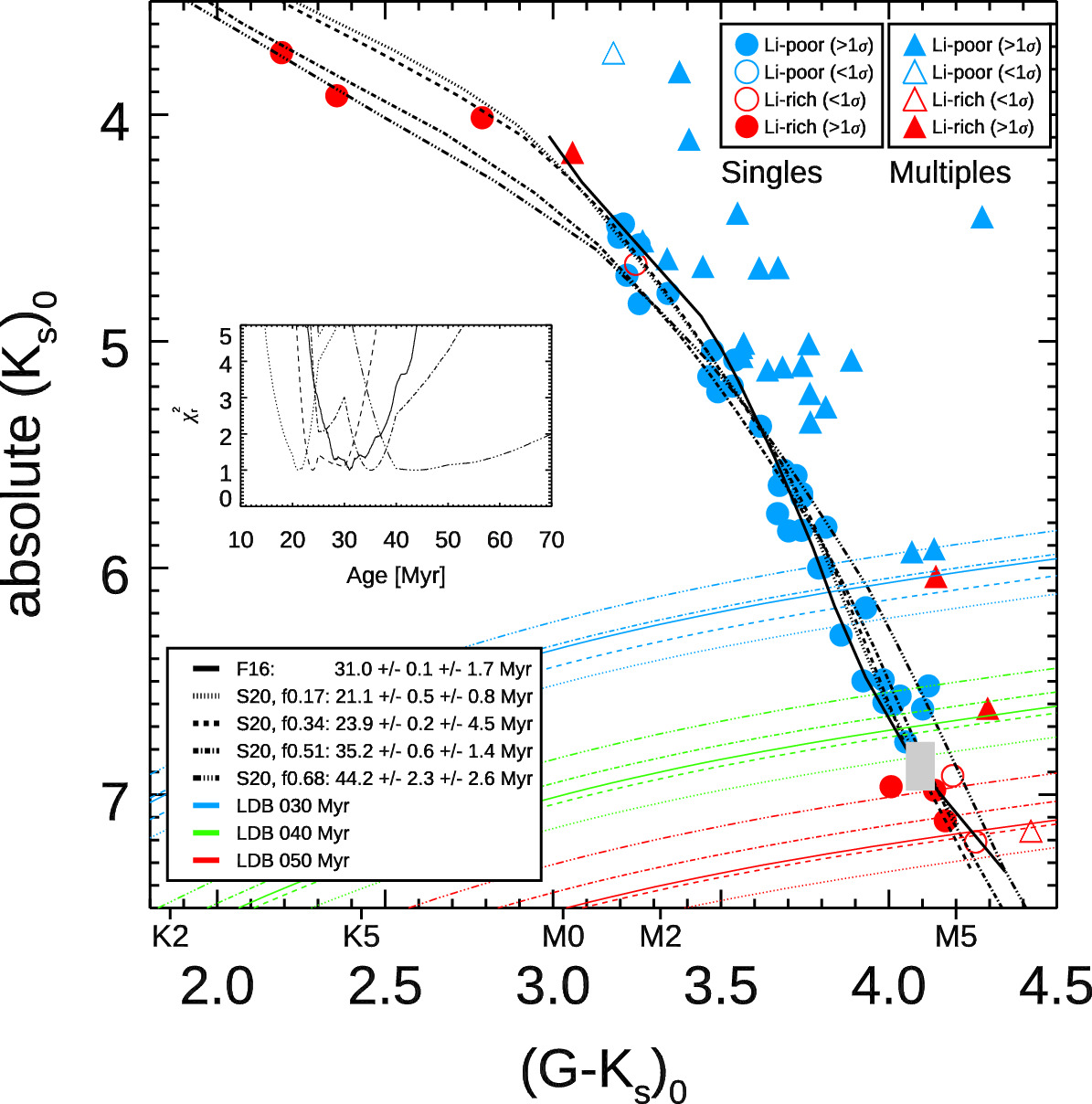}
\caption{Similar to Fig.~\ref{fig:CMD_SSM} but showing the loci of constant luminosity for 99 per cent Li depletion (coloured lines), the best fitting isochrones (black lines) and $\chi^2_r$ vs age for a series of magnetic models (see $\S$\ref{S_EvoModels}).}
\label{fig:CMD_NSSM}
\end{figure}

Bolometric luminosities and ages at 99 per cent Li-depletion were interpolated from the mass tracks of each model. The age sampling from the models was fine enough to follow the rapid Li-burning phase in fully convective stars. Figures~\ref{fig:CMD_SSM} and \ref{fig:CMD_NSSM} show example loci of the luminosity at 99 per cent Li-depletion at ages of 30, 40 and 50\,Myr, which encompass the observed LDB location (grey box). The loci are curved because the bolometric corrections are colour-dependent.

The LDB age is estimated as that of the 99 per cent depletion locus that passes through the centre of the LDB location. Age uncertainties are separated into a statistical and systematic component in Table~\ref{T_Age}. The statistical error bar is the quadrature sum due to the dimensions of the box defining the LDB location and corners the mean measurement error in $(G-K_{\rm s})_{0}$ and $M_{\rm K_{\rm s}}$ of the two targets that sit either side of the box. The systematic error is calculated as the quadrature sum due to shifting the LDB box (together with all the data points) by the uncertainties in distance and reddening/extinction. A shift of 0.1 mag in the absolute magnitude of the LDB box corresponds to a shift in age of $\sim 1$\,Myr. Horizontal shifts in the colour of the LDB are less important because the LDB luminosity loci are nearly horizontal. The overall age uncertainty is dominated by the vertical extent of the LDB box, which contributes $\sim 80$ per cent of the error budget.  We return to the effects of statistical and systematic errors on the LDB age in $\S$\ref{S_Discussion}.

The LDB ages determined using the three standard models are shown in Table~\ref{T_Age} and are in close agreement ($36.9-39.0$\,Myr). The choice of standard model contributes about the same uncertainty ($\sim 5$ per cent) as the total experimental uncertainties. This is unsurprising given that the physics in the interiors of fully-convective stars is relatively well understood and the known insensitivity of the LDB to the physical differences in models \citep{2004a_Burke, 2015a_Tognelli}. The analysis was repeated using the absolute $(G-J)_0/M_J$ diagram and the results are practically identical (see Table~\ref{T_Age}).

The magnetic models predict slightly older ages, by $\sim 5-20$ per cent compared with their standard model counterparts. The S20 models indicate that increasing magnetic activity (or at least a larger $f_{\rm sp}$) leads to older LDB ages. The results for the F16 model are similar to the S20 $f_{\rm sp}=0.34$ model. Again, the results using the $(G-J)_0/M_J$ diagram are almost identical.

There are enough data points surrounding the LDB box in Figures~\ref{fig:CMD_SSM} and~\ref{fig:CMD_NSSM} that we are reasonably confident the results are robust. In summary, the LDB ages determined from standard models are $37-39\,$Myr and from magnetic models are $41-50$\,Myr, and higher spot coverage/magnetic activity leads to the older ages.

{\centering
\begin{table*}
\begin{center}
\begin{tabular}{lrrrr}
\hline
\hline

Model          & $(G-K_{\rm s})_{\rm LDB}$ & $(G-J)_{\rm LDB}$ & $(G-K_{\rm s})_{\rm CMD}$ & $(G-J)_{\rm CMD}$ \\
\hline
\multicolumn{5}{c}{Standard models} \\
\hline
B15            & $36.9 \pm 1.8 \pm 0.7$ & $37.2 \pm 1.8 \pm 0.7$ & $16.0 \pm 0.4 \pm 1.2, 37, 0.039$ & $15.2 \pm 0.5 \pm 0.9, 34, 0.038$ \\
D08            & $37.5 \pm 1.9 \pm 0.8$ & $37.8 \pm 2.0 \pm 0.8$ & $17.7 \pm 0.5 \pm 1.1, 35, 0.034$ & $16.1 \pm 0.6 \pm 1.2, 32, 0.039$ \\
S20 (no spots) & $39.0 \pm 2.0 \pm 0.8$ & $39.3 \pm 2.1 \pm 0.9$ & $16.9 \pm 0.5 \pm 1.0, 43, 0.036$ & $15.6 \pm 0.7 \pm 1.1, 42, 0.053$ \\
\hline
\multicolumn{5}{c}{Non-standard models} \\
\hline
F16            & $43.2 \pm 2.1 \pm 0.8$ & $43.5 \pm 2.2 \pm 0.9$ & $31.0 \pm 0.1 \pm 1.7, 42, 0.033$ & $31.0 \pm 0.9 \pm 1.1, 38, 0.044$ \\
S20, f0.17     & $40.8 \pm 2.0 \pm 0.8$ & $41.0 \pm 2.1 \pm 0.9$ & $21.1 \pm 0.5 \pm 0.8, 42, 0.023$ & $20.7 \pm 0.2 \pm 0.6, 40, 0.037$ \\
S20, f0.34     & $42.7 \pm 2.1 \pm 0.9$ & $43.0 \pm 2.2 \pm 0.9$ & $23.9 \pm 0.2 \pm 4.5, 43, 0.026$ & $23.5 \pm 0.2 \pm 3.5, 40, 0.041$ \\
S20, f0.51     & $44.8 \pm 2.2 \pm 0.9$ & $45.1 \pm 2.3 \pm 1.0$ & $35.2 \pm 0.6 \pm 1.4, 43, 0.024$ & $34.0 \pm 0.4 \pm 1.2, 42, 0.041$ \\
S20, f0.68     & $47.3 \pm 2.5 \pm 1.0$ & $47.7 \pm 2.6 \pm 1.0$ & $44.2 \pm 2.3 \pm 2.6, 43, 0.052$ & $44.5 \pm 1.6 \pm 2.7, 42, 0.056$ \\
S20, f0.85     & $50.6 \pm 2.7 \pm 1.1$ & $50.2 \pm 2.7 \pm 1.1$ & -- & -- \\
\hline
\end{tabular}
\caption{Ages (in Myr) derived from the LDB method ($\S$\ref{S_LDB}) and from fitting isochrones to the CMD ($\S$\ref{S_CMD}). The first and second error bar are representative of the statistical and systematic uncertainties, respectively. For the CMD ages, the subsequent values are the number of targets that are used in the fitting process ($N$) and the figure-of-merit for the fit ($\xi$, described in $\S$\ref{S_CMD}), respectively. The upper section shows the results for standard models, the lower for magnetic models (see $\S$\ref{S_EvoModels}); B15 $=$ \protect\cite{2015a_Baraffe}; D08 $=$ \protect\cite{2008a_Dotter}; F16 $=$ \protect\cite{2016a_Feiden}; S20 $=$ \protect\cite{2020a_Somers}. No sensible fit to the CMD could be obtained for the S20 f0.85 model.}
\label{T_Age}
\end{center}
\end{table*}}

\section{Isochronal ages}\label{S_CMD}

Ages have also been estimated by fitting isochrones to the low-mass stars in absolute CMDs. Comparison is made in the observational plane since it makes the role of measurement uncertainties and systematic errors clear. Although a single $d_{\rm mod}$ value was adopted, cluster members will be at slightly different distances. The angular extent of the cluster members translates to a diameter of $\sim 6$\,pc. A similar front-to-back depth is too small to be resolved by the EDR3 parallaxes, so will introduce a modest additional scatter of $\sim \pm 0.02$ mag to the absolute magnitudes of individual stars. 

Cluster members will include a fraction of multiple systems that are brighter than single stars of the same colour. Including these in the isochrone fitting would bias the fits towards younger ages. Therefore we define a faint subsample that are likely to be single stars. These are cluster members with $M_{K_{\rm s}}$ greater than the interpolated value calculated from a second-order polynomial fit to $M_{K_{\rm s}}$ versus $(G-K_{\rm s})_0$ for the full target sample. Finally, we only retain targets for fitting that have $(G-K_{\rm s})_0>1.9$, since hotter and more luminous stars are likely to have reached the main sequence and will simply add noise to the fit. This leaves 43 ``faint'' members, likely to be single stars, which are denoted by circles in Figures~\ref{fig:CMD_SSM} and \ref{fig:CMD_NSSM} and 28 ``bright'' members (triangles), which are likely binary (or higher order multiple) systems. The spectral-types corresponding to the $(G-K_{\rm s})_0$ axes are from the main-sequence interpolation table provided by E. Mamajek\footnote{\url{https://www.pas.rochester.edu/~emamajek/EEM_dwarf_UBVIJHK_colors_Teff.txt}}. 

To estimate isochronal ages, a fit to the single stars is made using the same model isochrones and bolometric corrections described in $\S$\ref{S_EvoModels} at fixed values of $d_{\rm mod}$ and $E(B-V)$. The best-fit age minimises $\Delta^2/(N-1)$, where $\Delta^2$ is the sum of the squared residuals in $M_{K_{\rm s}}$  between the data and model isochrones and $N$ is the number of targets within the colour range covered by the model isochrone. 

This quantity provides a figure-of-merit that can be compared across different models. This method assumes that individual $M_{K_{\rm s}}$ uncertainties are homoscedastic and accounts for the fact that the dispersion around the best fit is large compared with the formal uncertainties in $M_{K_{\rm s}}$ from the photometry and adopted uniform distance -- probably because of stellar variability, rotational modulation and flares, the presence of some unresolved low mass-ratio binaries and perhaps as a result of varying spot coverage. The minimum value of $\Delta^2/(N-1)$ for each model is reported in Table~\ref{T_Age} along with the best-fitting isochrone.

To estimate statistical uncertainties in the ages, $\Delta^2/(N-1)$ is converted to a $\chi^2_r$-like statistic by normalising the $\Delta^2$ values so that the best-fitting $\Delta^2_{\rm min}/(N-1) = 1$. Uncertainties are then estimated from the ages at which the normalised $\Delta^2 = N$. This is equivalent to finding the ages such that $\chi^{2}=(N-1)\chi^{2}_{r,{\rm min}}+1$. The statistical age uncertainties are small ($\leq 0.6$\,Myr for all fits except the spottiest model). 

Systematic age uncertainties arise from the assumed values of $d_{\rm mod}= 7.518 \pm 0.021$ and $E(B-V)=0.07 \pm 0.02$. A larger distance and smaller reddening lead to younger ages. Since uncertainties in $d_{\rm mod}$ (from parallax) and $E(B-V)$ should be uncorrelated, then an additional systematic uncertainty was estimated from the quadrature sum of offsets in fitted age caused by changing $d_{\rm mod}$ and $E(B-V)$ (and the extinction) by their error bars. The best-fitting ages with both statistical and systematic error bars are reported in Table~\ref{T_Age}. 

\subsection{Standard models}\label{S_SSM}

 Figure~\ref{fig:CMD_SSM} shows the $M_{K_{\rm s}}$ versus $(G-K_{\rm s})_0$ CMD, with the best-fit isochrones from the 3 standard models. The best-fitting ages are given in the upper part of Table~\ref{T_Age}. The estimated ages are in close agreement; the age range of $16.0-17.7$\,Myr is comparable to the error bars. Ages from the $M_{J}$ versus $(G-J)_0$ are about $1$ Myr younger.

The standard models appear to be poor fits overall. They are all systematically overluminous by $0.1-0.4$\,mag at the hot end of the dataset and underluminous by $\sim 0.2$\,mag at the lowest temperatures. 

In contrast to the LDB ages, many stars are used in the fits to define an isochronal age and therefore the statistical precision of the ages is comparable to or better than the systematic errors attributable to distance and reddening uncertainty. An additional systematic uncertainty in the case of the isochrone fitting is the cut that was applied to define binary stars that were excluded from the fit. The default cut excluded 39 per cent of the members as binaries. The binary fraction among low-mass stars is unlikely to be larger than this and could well be smaller. We tested the impact of this by offsetting the threshold curve in the CMD that defined binarity until only 25 per cent of stars were considered binaries. The resulting isochronal fits were younger by just 1\,Myr and so we do not consider this any further as a significant source of age uncertainty.

\subsection{Magnetic models}

Figure~\ref{fig:CMD_NSSM} shows the results of fitting the magnetic model isochrones to the same data. The results are given in the lower part of Table~\ref{T_Age}. Two features are immediately clear: the magnetic models predict much older isochronal ages and a larger spread in age (from $21-44$ Myr) than the standard models. The spread is due to the range of magnetic activity considered. The youngest ages are for the least-spotted S20 models and the oldest are for the most-spotted models. The F16 magnetic models yield an age somewhere between the $f_{\rm sp}=0.34$ and $f_{\rm sp}=0.51$ models of S20. The $\chi^2$ landscape of the fits to the magnetic models is more complex than for the standard models and in some cases results in larger uncertainties, both statistical and systematic (particularly for the $f_{\rm sp}=0.34$ model).

The average absolute gradient of the S20 models become smaller with increasing $f_{\rm sp}$. The net effect of increasing the star-spot coverage in cooler stars is to cause inflated radii, lower $T_{\rm eff}$ and a higher luminosity at a given age. For higher-mass stars, with radiative cores, the radius inflation is much less and the luminosity is reduced. The S20 models with $0.17\leq f_{\rm sp}\leq 0.51$ provide the best fits to the data according to the unnormalised $\Delta^2/(N-1)$ statistic and they appear to fit the data better than any of the standard models. The significance of this can be assessed using the difference in $N\ln\Delta^2$, equivalent to the Akaike information criterion in least squares fitting. For example, this difference is 17.44 between the $f_{\rm sp}=0.51$ and the equivalent unspotted S20 models; a likelihood ratio of $\exp(-17.44/2)=1.6\times 10^{-4}$ in favour of the spotted model.

A more spotted $f_{\rm sp}=0.68$ model however provides a significantly worse fit. It matches the high-mass stars reasonably, but becomes increasingly overluminous redward of $(G-K_{\rm s})=3.5$. We were unable to find any reasonable fit for a $f_{\rm sp}=0.85$ model and discarded these from our analysis. The F16 models provide a reasonable fit at low masses, but they do not extend to colours blueward of $(G-K_{\rm s})=3.0$ where they appear overluminous by $\sim 0.2$ mag.

\section{Discussion}\label{S_Discussion}

\begin{figure}
\centering
  \includegraphics[width=.48\textwidth]{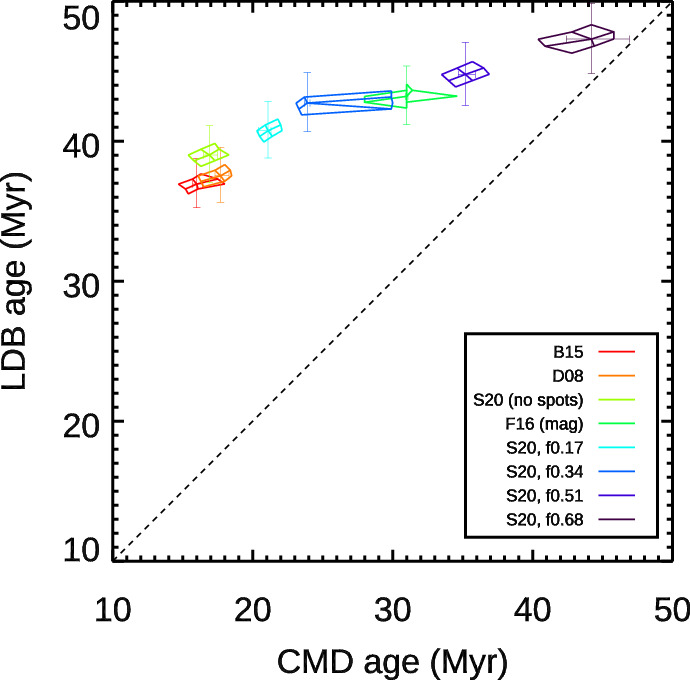}
\caption{A comparison of the LDB- and CMD-derived ages from each model. Vertical and horizontal error bars represent the age error due to statistical uncertainty and the grids demonstrate the effect on the derived age when altering $E(B-V)$ and $d_{\rm mod}$ by their error bars (representing systematic uncertainties). The dotted line denotes equivalence.}
\label{fig:age_comp}
\end{figure}

Figure~\ref{fig:age_comp} compares the ages estimated from the LDB and CMD method for each model. The errors due to statistical and systematic uncertainty are indicated. Statistical errors dominate for the LDB ages because the LDB location is determined by only a handful of stars. The effects of systematic errors due to uncertainties in distance and reddening are small. In contrast, many more data points define the best-fit CMD ages and the effects of distance and reddening uncertainties are comparable or larger than the statistical errors.

The LDB ages are older than the CMD ages for all the models tested; much older (by factors of $\geq 2$) for standard models, but becoming closer, but with older ages, for models featuring increasing levels of magnetic activity, to the extent that there is marginal consistency for the $f_{\rm sp}=0.68$ model of S20. There is excellent internal agreement between the standard models in terms of what they give for the LDB ages, the CMD ages, and their ratio, despite the differing ingredients in these models in terms of atmospheres, convection and boundary conditions. There are much larger differences between the magnetic models, but that is expected since they represent differing levels of magnetic activity. What is notable about Fig.~\ref{fig:age_comp} is how little the LDB ages increase with increased magnetic activity compared to the very large increases in CMD age. In other words, the LDB ages are {\it much} less sensitive to the effects of magnetic activity, and even less so to other varying model inputs \citep[see also][]{2004a_Burke, 2014b_Jackson, 2015a_Tognelli}, and should therefore be given more weight.

That the magnetic models predict CMD ages for NGC 2232 up to 2-3 times older than from standard models is in broad agreement with \cite{2013a_Bell}, who found that low-mass isochronal ages needed to be significantly increased to match ages found from isochronal fits to high-mass members of the same young clusters.
It also agrees with work that finds the isochronal ages and the ages inferred from the general pattern of Li-depletion among low-mass members of young clusters do not agree when using standard models, but might be brought into agreement at significantly older ages if the low-mass stars are inflated by magnetic activity and starspots  \citep{2015a_Somers, 2017a_Jeffries}. The turn-off age of NGC~2232 is rather uncertain and based on few stars, $32\pm 15$\,Myr using models with some core overshooting \citep{2006a_Lyra}, and might agree with any of the ages in this work.

It appears that very high levels of magnetic activity or spot coverage ($f_{\rm sp} \geq 0.68$) might be required, in order to bring isochronal and LDB ages into agreement at $\geq 45$ Myr, although the very high spot coverage models are not a good fit to the entire low-mass sequence in NGC 2232. There is however no compelling reason why spot coverage or magnetic activity should be uniform with stellar mass. It could be that the lower mass stars have much higher levels of spot coverage for example. A ``hybrid model'', featuring higher spot coverage at lower masses, with fewer spots at higher masses or models with cooler spots could provide an excellent fit to the CMD. Zeeman Doppler Imaging studies do suggest that the magnetic field strength in the higher mass stars is weaker than in the low-mass stars (see, e.g., figure 10 in \citealt{2018a_Folsom} and figure 14 in \citealt{2021a_Kochukhov}).

Is a $>50$ per cent spot covering fraction realistic for the low mass stars in NGC~2232? M-dwarf members of young clusters are generally found to be both rapidly rotating and highly magnetically active. The M-dwarfs in NGC~2547, which has a similar LDB age of $35\pm 3$\,Myr \citep[based on standard models,][]{2005a_Jeffries} have saturated levels of coronal X-ray activity \citep{2011a_Jeffries}. It has been verified that highly-active M-dwarfs have strong magnetism, consistent with equipartition magnetic fields covering all of their surfaces \citep{2009a_Reiners, 2010a_Morin}. \cite{2009a_Jackson} find that a high spot coverage ($>50$ per cent) is needed to fit the CMD for M-dwarfs in NGC2516 (age $\sim 120$ Myr); \cite{2016a_Fang} estimated $f_{\rm sp}$ values of up to 50 per cent for Pleiades K and M dwarfs at $\sim 100$ Myr by modelling their molecular absorption bands, and \cite{2018a_Jackson} found evidence that these M dwarfs were correspondingly larger than predicted by standard evolutionary models and in better accord with the magnetic models.

\section{Summary}\label{S_Summary}

A combination of astrometry from the Gaia mission and EW(Li) observations from GES has been used to select a high probability set of low-mass members in NGC~2232. We have identified the lithium depletion boundary (LDB) - the luminosity at which low mass stars switch from having undepleted photospheric Li to having no detectable Li at only slightly higher luminosities. The LDB is used to determine the age of the cluster with standard evolutionary models and also with ``magnetic models" that incorporate magnetic suppression of convection or the blocking of flux by dark, magnetic  starspots at the surface. These ages are compared with those determined by fitting isochrones to low-mass stars in colour-magnitude diagrams using the same models. The results and conclusions are:

\begin{itemize}
  \item The LDB age of NGC~2232 is $38 \pm 3$ Myr using standard models. The uncertainty is dominated by locating the LDB, which is defined by only a few stars. Systematic errors due to the adopted distance and reddening, or choice of standard model, are less important.

\item LDB ages using the magnetic models are between 41 Myr and 50 Myr, with similar error bars. The inferred age increases with the fraction of the stellar photospheres assumed to be covered with dark spots.

\item Isochronal ages from fitting the CMD are more than a factor of two younger (15--18\,Myr) than the LDB ages in the case of the standard models, but can be much closer (20--44\,Myr) for magnetic models, with agreement between the separate techniques being best for models with spot coverage fractions $>50$ per cent.
\end{itemize}
The LDB ages are much more robust to the input physics and the levels of magnetic activity assumed. The strong disagreement between the LDB ages and ages derived from isochrone fits using the standard models indicates that there is missing physics in those models. The much better agreement between the LDB and isochronal ages for the magnetic models suggests that these offer a significantly better description of the early evolution of low-mass stars. If so, then ages determined by fitting isochrones to young, low-mass stars using standard models will need to be revised upwards by factors of 2-3, and masses inferred from standard models and the positions of low-mass stars in the CMD or HR diagram will be underestimates.

\section{Acknowledgements}\label{S_Ack}
ASB, RDJ and RJJ acknowledge the financial support of the STFC. We are grateful to I. Baraffe for providing a version of the B15 models with finer age sampling which was used in our analysis of the lithium depletion boundary.

Based on data products from observations made with ESO Telescopes at the La Silla Paranal Observatory under programme ID 188.B-3002. These data products have been processed by the Cambridge Astronomy Survey Unit (CASU) at the Institute of Astronomy, University of Cambridge, and by the FLAMES/UVES reduction team at INAF/Osservatorio Astrofisico di Arcetri. These data have been obtained from the Gaia-ESO Survey Data Archive, prepared and hosted by the Wide Field Astronomy Unit, Institute for Astronomy, University of Edinburgh, which is funded by the UK Science and Technology Facilities Council. This work was partly supported by the European Union FP7 programme through ERC grant number 320360 and by the Leverhulme Trust through grant RPG-2012-541. We acknowledge the support from INAF and Ministero dell’ Istruzione, dell’ Universit\`a e della Ricerca (MIUR) in the form of the grant ``Premiale VLT 2012''. The results presented here benefit from discussions held during the Gaia-ESO workshops and conferences supported by the ESF (European Science Foundation) through the GREAT Research Network Programme. This work has made use of data from the European Space Agency (ESA) mission Gaia \url{https://www.cosmos.esa.int/gaia}), processed by the Gaia Data Processing and Analysis Consortium (DPAC, \url{https://www.cosmos.esa.int/web/gaia/dpac/consortium}). Funding for the DPAC has been provided by national institutions, in particular the institutions participating in the Gaia Multilateral Agreement.

\section{Data Availability Statement}\label{S_DataAvailability}

The quantities used to perform the analyses in $\S$\ref{S_LDB}~and $\S$\ref{S_CMD} are presented in Table~\ref{T_Data}. All the data are available from the ESO archive in the form of raw and reduced spectra\footnote{\url{https://archive.eso.org/cms/eso-archive-news/New-data-release-DR4-from-the-Gaia-ESO-Spectroscopic-Public-Survey.html}}.

\bibliographystyle{mnras}
\bibliography{bibliography}

\begin{thebibliography}{}
\makeatletter
\relax
\def\mn@urlcharsother{\let\do\@makeother \do\$\do\&\do\#\do\^\do\_\do\%\do\~}
\def\mn@doi{\begingroup\mn@urlcharsother \@ifnextchar [ {\mn@doi@}
  {\mn@doi@[]}}
\def\mn@doi@[#1]#2{\def\@tempa{#1}\ifx\@tempa\@empty \href
  {http://dx.doi.org/#2} {doi:#2}\else \href {http://dx.doi.org/#2} {#1}\fi
  \endgroup}
\def\mn@eprint#1#2{\mn@eprint@#1:#2::\@nil}
\def\mn@eprint@arXiv#1{\href {http://arxiv.org/abs/#1} {{\tt arXiv:#1}}}
\def\mn@eprint@dblp#1{\href {http://dblp.uni-trier.de/rec/bibtex/#1.xml}
  {dblp:#1}}
\def\mn@eprint@#1:#2:#3:#4\@nil{\def\@tempa {#1}\def\@tempb {#2}\def\@tempc
  {#3}\ifx \@tempc \@empty \let \@tempc \@tempb \let \@tempb \@tempa \fi \ifx
  \@tempb \@empty \def\@tempb {arXiv}\fi \@ifundefined
  {mn@eprint@\@tempb}{\@tempb:\@tempc}{\expandafter \expandafter \csname
  mn@eprint@\@tempb\endcsname \expandafter{\@tempc}}}

\bibitem[\protect\citeauthoryear{{Baraffe}, {Homeier}, {Allard}  \&
  {Chabrier}}{{Baraffe} et~al.}{2015}]{2015a_Baraffe}
{Baraffe} I.,  {Homeier} D.,  {Allard} F.,   {Chabrier} G.,  2015, \mn@doi
  [\aap] {10.1051/0004-6361/201425481}, \href
  {http://adsabs.harvard.edu/abs/2015A\%26A...577A..42B} {577, A42}

\bibitem[\protect\citeauthoryear{{Beccari} et~al.,}{{Beccari}
  et~al.}{2017}]{2017a_Beccari}
{Beccari} G.,  et~al., 2017, \mn@doi [\aap] {10.1051/0004-6361/201730432},
  \href {https://ui.adsabs.harvard.edu/abs/2017A&A...604A..22B} {604, A22}

\bibitem[\protect\citeauthoryear{{Bell}, {Naylor}, {Mayne}, {Jeffries}  \&
  {Littlefair}}{{Bell} et~al.}{2013}]{2013a_Bell}
{Bell} C.~P.~M.,  {Naylor} T.,  {Mayne} N.~J.,  {Jeffries} R.~D.,
  {Littlefair} S.~P.,  2013, \mn@doi [\mnras] {10.1093/mnras/stt1075}, \href
  {http://adsabs.harvard.edu/abs/2013MNRAS.434..806B} {434, 806}

\bibitem[\protect\citeauthoryear{{Bildsten}, {Brown}, {Matzner}  \&
  {Ushomirsky}}{{Bildsten} et~al.}{1997}]{1997a_Bildsten}
{Bildsten} L.,  {Brown} E.~F.,  {Matzner} C.~D.,   {Ushomirsky} G.,  1997,
  \mn@doi [\apj] {10.1086/304151}, \href
  {http://adsabs.harvard.edu/abs/1997ApJ...482..442B} {482, 442}

\bibitem[\protect\citeauthoryear{{Binks} \& {Jeffries}}{{Binks} \&
  {Jeffries}}{2014}]{2014a_Binks}
{Binks} A.~S.,  {Jeffries} R.~D.,  2014, \mn@doi [\mnras]
  {10.1093/mnrasl/slt141}, \href
  {http://adsabs.harvard.edu/abs/2014MNRAS.438L..11B} {438, L11}

\bibitem[\protect\citeauthoryear{{Binks} \& {Jeffries}}{{Binks} \&
  {Jeffries}}{2016}]{2016a_Binks}
{Binks} A.~S.,  {Jeffries} R.~D.,  2016, \mn@doi [\mnras]
  {10.1093/mnras/stv2431}, \href
  {http://adsabs.harvard.edu/abs/2016MNRAS.455.3345B} {455, 3345}

\bibitem[\protect\citeauthoryear{{Bouvier} et~al.,}{{Bouvier}
  et~al.}{2016}]{2016a_Bouvier}
{Bouvier} J.,  et~al., 2016, \mn@doi [\aap] {10.1051/0004-6361/201628336},
  \href {https://ui.adsabs.harvard.edu/abs/2016A&A...590A..78B} {590, A78}

\bibitem[\protect\citeauthoryear{{Bouvier} et~al.,}{{Bouvier}
  et~al.}{2018}]{2018a_Bouvier}
{Bouvier} J.,  et~al., 2018, \mn@doi [\aap] {10.1051/0004-6361/201731881},
  \href {http://adsabs.harvard.edu/abs/2018A%26A...613A..63B} {613, A63}

\bibitem[\protect\citeauthoryear{{Burke}, {Pinsonneault}  \& {Sills}}{{Burke}
  et~al.}{2004}]{2004a_Burke}
{Burke} C.~J.,  {Pinsonneault} M.~H.,   {Sills} A.,  2004, \mn@doi [\apj]
  {10.1086/381242}, \href {http://adsabs.harvard.edu/abs/2004ApJ...604..272B}
  {604, 272}

\bibitem[\protect\citeauthoryear{{Cardelli}, {Clayton}  \& {Mathis}}{{Cardelli}
  et~al.}{1989}]{1989a_Cardelli}
{Cardelli} J.~A.,  {Clayton} G.~C.,   {Mathis} J.~S.,  1989, \mn@doi [\apj]
  {10.1086/167900}, \href
  {https://ui.adsabs.harvard.edu/abs/1989ApJ...345..245C} {345, 245}

\bibitem[\protect\citeauthoryear{{Casali} et~al.,}{{Casali}
  et~al.}{2019}]{2019a_Casali}
{Casali} G.,  et~al., 2019, \mn@doi [\aap] {10.1051/0004-6361/201935282}, \href
  {https://ui.adsabs.harvard.edu/abs/2019A&A...629A..62C} {629, A62}

\bibitem[\protect\citeauthoryear{{Casali} et~al.,}{{Casali}
  et~al.}{2020}]{2020a_Casali}
{Casali} G.,  et~al., 2020, \mn@doi [\aap] {10.1051/0004-6361/202038055}, \href
  {https://ui.adsabs.harvard.edu/abs/2020A&A...639A.127C} {639, A127}

\bibitem[\protect\citeauthoryear{{Claria}}{{Claria}}{1972}]{1972a_Claria}
{Claria} J.~J.,  1972, \aap, \href
  {https://ui.adsabs.harvard.edu/abs/1972A&A....19..303C} {19, 303}

\bibitem[\protect\citeauthoryear{{Collinder}}{{Collinder}}{1931}]{1931a_Collinder}
{Collinder} P.,  1931, Annals of the Observatory of Lund, \href
  {https://ui.adsabs.harvard.edu/abs/1931AnLun...2....1C} {2, B1}

\bibitem[\protect\citeauthoryear{{Cutri} et~al.,}{{Cutri}
  et~al.}{2003}]{2003a_Cutri}
{Cutri} R.~M.,  et~al., 2003, {2MASS All Sky Catalog of point sources.}

\bibitem[\protect\citeauthoryear{{Damiani} et~al.,}{{Damiani}
  et~al.}{2014}]{2014a_Damiani}
{Damiani} F.,  et~al., 2014, \mn@doi [\aap] {10.1051/0004-6361/201323306},
  \href {https://ui.adsabs.harvard.edu/abs/2014A&A...566A..50D} {566, A50}

\bibitem[\protect\citeauthoryear{{Danielski}, {Babusiaux}, {Ruiz-Dern},
  {Sartoretti}  \& {Arenou}}{{Danielski} et~al.}{2018}]{2018a_Danielski}
{Danielski} C.,  {Babusiaux} C.,  {Ruiz-Dern} L.,  {Sartoretti} P.,   {Arenou}
  F.,  2018, \mn@doi [\aap] {10.1051/0004-6361/201732327}, \href
  {https://ui.adsabs.harvard.edu/abs/2018A&A...614A..19D} {614, A19}

\bibitem[\protect\citeauthoryear{{De Marchi}, {Panagia}, {Guarcello}  \&
  {Bonito}}{{De Marchi} et~al.}{2013}]{2013a_De_Marchi}
{De Marchi} G.,  {Panagia} N.,  {Guarcello} M.~G.,   {Bonito} R.,  2013,
  \mn@doi [\mnras] {10.1093/mnras/stt1499}, \href
  {https://ui.adsabs.harvard.edu/abs/2013MNRAS.435.3058D} {435, 3058}

\bibitem[\protect\citeauthoryear{{Dotter}, {Chaboyer}, {Jevremovi{\'c}},
  {Kostov}, {Baron}  \& {Ferguson}}{{Dotter} et~al.}{2008}]{2008a_Dotter}
{Dotter} A.,  {Chaboyer} B.,  {Jevremovi{\'c}} D.,  {Kostov} V.,  {Baron} E.,
  {Ferguson} J.~W.,  2008, \mn@doi [\apjs] {10.1086/589654}, \href
  {http://adsabs.harvard.edu/abs/2008ApJS..178...89D} {178, 89}

\bibitem[\protect\citeauthoryear{{Dreyer}}{{Dreyer}}{1888}]{1888a_Dreyer}
{Dreyer} J.~L.~E.,  1888, \memras, \href
  {https://ui.adsabs.harvard.edu/abs/1888MmRAS..49....1D} {49, 1}

\bibitem[\protect\citeauthoryear{{Fang}, {Zhao}, {Zhao}, {Chen}  \& {Bharat
  Kumar}}{{Fang} et~al.}{2016}]{2016a_Fang}
{Fang} X.-S.,  {Zhao} G.,  {Zhao} J.-K.,  {Chen} Y.-Q.,   {Bharat Kumar} Y.,
  2016, \mn@doi [\mnras] {10.1093/mnras/stw1923}, \href
  {https://ui.adsabs.harvard.edu/abs/2016MNRAS.463.2494F} {463, 2494}

\bibitem[\protect\citeauthoryear{{Feiden}}{{Feiden}}{2016}]{2016a_Feiden}
{Feiden} G.~A.,  2016, \mn@doi [\aap] {10.1051/0004-6361/201527613}, \href
  {http://adsabs.harvard.edu/abs/2016A\%26A...593A..99F} {593, A99}

\bibitem[\protect\citeauthoryear{{Feiden} \& {Chaboyer}}{{Feiden} \&
  {Chaboyer}}{2013}]{2013a_Feiden}
{Feiden} G.~A.,  {Chaboyer} B.,  2013, \mn@doi [\apj]
  {10.1088/0004-637X/779/2/183}, \href
  {http://adsabs.harvard.edu/abs/2013ApJ...779..183F} {779, 183}

\bibitem[\protect\citeauthoryear{{Feiden} \& {Chaboyer}}{{Feiden} \&
  {Chaboyer}}{2014}]{2014a_Feiden}
{Feiden} G.~A.,  {Chaboyer} B.,  2014, \mn@doi [\apj]
  {10.1088/0004-637X/789/1/53}, \href
  {http://adsabs.harvard.edu/abs/2014ApJ...789...53F} {789, 53}

\bibitem[\protect\citeauthoryear{{Folsom} et~al.,}{{Folsom}
  et~al.}{2018}]{2018a_Folsom}
{Folsom} C.~P.,  et~al., 2018, \mn@doi [\mnras] {10.1093/mnras/stx3021}, \href
  {http://adsabs.harvard.edu/abs/2018MNRAS.474.4956F} {474, 4956}

\bibitem[\protect\citeauthoryear{{Gaia Collaboration} et~al.,}{{Gaia
  Collaboration} et~al.}{2018}]{2018a_Gaia_Collaboration}
{Gaia Collaboration} et~al., 2018, \mn@doi [\aap]
  {10.1051/0004-6361/201833051}, \href
  {https://ui.adsabs.harvard.edu/abs/2018A&A...616A...1G} {616, A1}

\bibitem[\protect\citeauthoryear{{Gaia Collaboration}, {Brown}, {Vallenari},
  {Prusti}, {de Bruijne}, {Babusiaux}  \& {Biermann}}{{Gaia Collaboration}
  et~al.}{2020}]{2020a_Gaia_Collaboration}
{Gaia Collaboration} {Brown} A.~G.~A.,  {Vallenari} A.,  {Prusti} T.,  {de
  Bruijne} J.~H.~J.,  {Babusiaux} C.,   {Biermann} M.,  2020, arXiv e-prints,
  \href {https://ui.adsabs.harvard.edu/abs/2020arXiv201201533G} {p.
  arXiv:2012.01533}

\bibitem[\protect\citeauthoryear{{Gilmore} et~al.,}{{Gilmore}
  et~al.}{2012}]{2012a_Gilmore}
{Gilmore} G.,  et~al., 2012, The Messenger, \href
  {http://adsabs.harvard.edu/abs/2012Msngr.147...25G} {147, 25}

\bibitem[\protect\citeauthoryear{{Gray}}{{Gray}}{1984}]{1984a_Gray}
{Gray} D.~F.,  1984, \mn@doi [\apj] {10.1086/161735}, \href
  {http://adsabs.harvard.edu/abs/1984ApJ...277..640G} {277, 640}

\bibitem[\protect\citeauthoryear{{Haisch}, {Lada}  \& {Lada}}{{Haisch}
  et~al.}{2001}]{2001a_Haisch}
{Haisch} Jr. K.~E.,  {Lada} E.~A.,   {Lada} C.~J.,  2001, \mn@doi [\apjl]
  {10.1086/320685}, \href {http://adsabs.harvard.edu/abs/2001ApJ...553L.153H}
  {553, L153}

\bibitem[\protect\citeauthoryear{{Herczeg} \& {Hillenbrand}}{{Herczeg} \&
  {Hillenbrand}}{2015}]{2015a_Herczeg}
{Herczeg} G.~J.,  {Hillenbrand} L.~A.,  2015, \mn@doi [\apj]
  {10.1088/0004-637X/808/1/23}, \href
  {http://adsabs.harvard.edu/abs/2015ApJ...808...23H} {808, 23}

\bibitem[\protect\citeauthoryear{{Herschel}}{{Herschel}}{1864}]{1864a_Herschel}
{Herschel} J. F.~W.,  1864, Philosophical Transactions of the Royal Society of
  London Series I, \href
  {https://ui.adsabs.harvard.edu/abs/1864RSPT..154....1H} {154, 1}

\bibitem[\protect\citeauthoryear{{Horne} \& {Baliunas}}{{Horne} \&
  {Baliunas}}{1986}]{1986a_Horne}
{Horne} J.~H.,  {Baliunas} S.~L.,  1986, \mn@doi [\apj] {10.1086/164037}, \href
  {http://adsabs.harvard.edu/abs/1986ApJ...302..757H} {302, 757}

\bibitem[\protect\citeauthoryear{{Jackson} \& {Jeffries}}{{Jackson} \&
  {Jeffries}}{2014a}]{2014a_Jackson}
{Jackson} R.~J.,  {Jeffries} R.~D.,  2014a, \mn@doi [\mnras]
  {10.1093/mnras/stu651}, \href
  {https://ui.adsabs.harvard.edu/abs/2014MNRAS.441.2111J} {441, 2111}

\bibitem[\protect\citeauthoryear{{Jackson} \& {Jeffries}}{{Jackson} \&
  {Jeffries}}{2014b}]{2014b_Jackson}
{Jackson} R.~J.,  {Jeffries} R.~D.,  2014b, \mn@doi [\mnras]
  {10.1093/mnras/stu2076}, \href
  {https://ui.adsabs.harvard.edu/abs/2014MNRAS.445.4306J} {445, 4306}

\bibitem[\protect\citeauthoryear{{Jackson}, {Jeffries}  \& {Maxted}}{{Jackson}
  et~al.}{2009}]{2009a_Jackson}
{Jackson} R.~J.,  {Jeffries} R.~D.,   {Maxted} P.~F.~L.,  2009, \mn@doi
  [\mnras] {10.1111/j.1745-3933.2009.00729.x}, \href
  {http://adsabs.harvard.edu/abs/2009MNRAS.399L..89J} {399, L89}

\bibitem[\protect\citeauthoryear{{Jackson} et~al.,}{{Jackson}
  et~al.}{2015}]{2015a_Jackson}
{Jackson} R.~J.,  et~al., 2015, \mn@doi [\aap] {10.1051/0004-6361/201526248},
  \href {https://ui.adsabs.harvard.edu/abs/2015A&A...580A..75J} {580, A75}

\bibitem[\protect\citeauthoryear{{Jackson}, {Deliyannis}  \&
  {Jeffries}}{{Jackson} et~al.}{2018}]{2018a_Jackson}
{Jackson} R.~J.,  {Deliyannis} C.~P.,   {Jeffries} R.~D.,  2018, \mn@doi
  [\mnras] {10.1093/mnras/sty374}, \href
  {http://adsabs.harvard.edu/abs/2018MNRAS.476.3245J} {476, 3245}

\bibitem[\protect\citeauthoryear{{Jackson} et~al.,}{{Jackson}
  et~al.}{2020}]{2020a_Jackson}
{Jackson} R.~J.,  et~al., 2020, \mn@doi [\mnras] {10.1093/mnras/staa1749},
  \href {https://ui.adsabs.harvard.edu/abs/2020MNRAS.tmp.1884J} {}

\bibitem[\protect\citeauthoryear{{Jeffries} \& {Oliveira}}{{Jeffries} \&
  {Oliveira}}{2005}]{2005a_Jeffries}
{Jeffries} R.~D.,  {Oliveira} J.~M.,  2005, \mn@doi [\mnras]
  {10.1111/j.1365-2966.2005.08820.x}, \href
  {http://adsabs.harvard.edu/abs/2005MNRAS.358...13J} {358, 13}

\bibitem[\protect\citeauthoryear{{Jeffries}, {Littlefair}, {Naylor}  \&
  {Mayne}}{{Jeffries} et~al.}{2011}]{2011a_Jeffries}
{Jeffries} R.~D.,  {Littlefair} S.~P.,  {Naylor} T.,   {Mayne} N.~J.,  2011,
  \mn@doi [\mnras] {10.1111/j.1365-2966.2011.19613.x}, \href
  {http://adsabs.harvard.edu/abs/2011MNRAS.418.1948J} {418, 1948}

\bibitem[\protect\citeauthoryear{{Jeffries}, {Naylor}, {Mayne}, {Bell}  \&
  {Littlefair}}{{Jeffries} et~al.}{2013}]{2013a_Jeffries}
{Jeffries} R.~D.,  {Naylor} T.,  {Mayne} N.~J.,  {Bell} C.~P.~M.,
  {Littlefair} S.~P.,  2013, \mn@doi [\mnras] {10.1093/mnras/stt1180}, \href
  {http://adsabs.harvard.edu/abs/2013MNRAS.434.2438J} {434, 2438}

\bibitem[\protect\citeauthoryear{{Jeffries} et~al.,}{{Jeffries}
  et~al.}{2014}]{2014b_Jeffries}
{Jeffries} R.~D.,  et~al., 2014, \mn@doi [\aap] {10.1051/0004-6361/201323288},
  \href {http://adsabs.harvard.edu/abs/2014A\%26A...563A..94J} {563, A94}

\bibitem[\protect\citeauthoryear{{Jeffries} et~al.,}{{Jeffries}
  et~al.}{2017}]{2017a_Jeffries}
{Jeffries} R.~D.,  et~al., 2017, \mn@doi [\mnras] {10.1093/mnras/stw2458},
  \href {http://adsabs.harvard.edu/abs/2017MNRAS.464.1456J} {464, 1456}

\bibitem[\protect\citeauthoryear{{Jofr{\'e}} et~al.,}{{Jofr{\'e}}
  et~al.}{2014}]{2014a_Jofre}
{Jofr{\'e}} P.,  et~al., 2014, \mn@doi [\aap] {10.1051/0004-6361/201322440},
  \href {https://ui.adsabs.harvard.edu/abs/2014A&A...564A.133J} {564, A133}

\bibitem[\protect\citeauthoryear{{Kochukhov}}{{Kochukhov}}{2021}]{2021a_Kochukhov}
{Kochukhov} O.,  2021, \mn@doi [\aapr] {10.1007/s00159-020-00130-3}, \href
  {https://ui.adsabs.harvard.edu/abs/2021A&ARv..29....1K} {29, 1}

\bibitem[\protect\citeauthoryear{{Kraus}, {Cody}, {Covey}, {Rizzuto}, {Mann}
  \& {Ireland}}{{Kraus} et~al.}{2015}]{2015a_Kraus}
{Kraus} A.~L.,  {Cody} A.~M.,  {Covey} K.~R.,  {Rizzuto} A.~C.,  {Mann} A.~W.,
   {Ireland} M.~J.,  2015, \mn@doi [\apj] {10.1088/0004-637X/807/1/3}, \href
  {https://ui.adsabs.harvard.edu/abs/2015ApJ...807....3K} {807, 3}

\bibitem[\protect\citeauthoryear{{Kraus} et~al.,}{{Kraus}
  et~al.}{2017}]{2017b_Kraus}
{Kraus} A.~L.,  et~al., 2017, \mn@doi [\apj] {10.3847/1538-4357/aa7e75}, \href
  {https://ui.adsabs.harvard.edu/abs/2017ApJ...845...72K} {845, 72}

\bibitem[\protect\citeauthoryear{{Levato} \& {Malaroda}}{{Levato} \&
  {Malaroda}}{1974}]{1974a_Levato}
{Levato} H.,  {Malaroda} S.,  1974, \mn@doi [\aj] {10.1086/111625}, \href
  {https://ui.adsabs.harvard.edu/abs/1974AJ.....79..890L} {79, 890}

\bibitem[\protect\citeauthoryear{{Lindegren} et~al.,}{{Lindegren}
  et~al.}{2018}]{2018a_Lindegren}
{Lindegren} L.,  et~al., 2018, \mn@doi [\aap] {10.1051/0004-6361/201832727},
  \href {http://adsabs.harvard.edu/abs/2018A%26A...616A...2L} {616, A2}

\bibitem[\protect\citeauthoryear{{Lindegren} et~al.,}{{Lindegren}
  et~al.}{2020}]{2020a_Lindegren}
{Lindegren} L.,  et~al., 2020, arXiv e-prints, \href
  {https://ui.adsabs.harvard.edu/abs/2020arXiv201201742L} {p. arXiv:2012.01742}

\bibitem[\protect\citeauthoryear{{Luhman}}{{Luhman}}{2007}]{2007a_Luhman}
{Luhman} K.~L.,  2007, \mn@doi [\apjs] {10.1086/520114}, \href
  {http://adsabs.harvard.edu/abs/2007ApJS..173..104L} {173, 104}

\bibitem[\protect\citeauthoryear{{Lyra}, {Moitinho}, {van der Bliek}  \&
  {Alves}}{{Lyra} et~al.}{2006}]{2006a_Lyra}
{Lyra} W.,  {Moitinho} A.,  {van der Bliek} N.~S.,   {Alves} J.,  2006, \mn@doi
  [\aap] {10.1051/0004-6361:20053894}, \href
  {http://adsabs.harvard.edu/abs/2006A\%26A...453..101L} {453, 101}

\bibitem[\protect\citeauthoryear{{MacDonald} \& {Mullan}}{{MacDonald} \&
  {Mullan}}{2017}]{2017a_Macdonald}
{MacDonald} J.,  {Mullan} D.~J.,  2017, \mn@doi [\apj]
  {10.3847/1538-4357/834/1/67}, \href
  {https://ui.adsabs.harvard.edu/abs/2017ApJ...834...67M} {834, 67}

\bibitem[\protect\citeauthoryear{{MacDonald} \& {Mullan}}{{MacDonald} \&
  {Mullan}}{2021}]{2021a_Macdonald}
{MacDonald} J.,  {Mullan} D.~J.,  2021, \mn@doi [\apj]
  {10.3847/1538-4357/abcfbf}, \href
  {https://ui.adsabs.harvard.edu/abs/2021ApJ...907...27M} {907, 27}

\bibitem[\protect\citeauthoryear{{Malo}, {Artigau}, {Doyon}, {Lafreni{\`e}re},
  {Albert}  \& {Gagn{\'e}}}{{Malo} et~al.}{2014a}]{2014a_Malo}
{Malo} L.,  {Artigau} {\'E}.,  {Doyon} R.,  {Lafreni{\`e}re} D.,  {Albert} L.,
   {Gagn{\'e}} J.,  2014a, \mn@doi [\apj] {10.1088/0004-637X/788/1/81}, \href
  {http://adsabs.harvard.edu/abs/2014ApJ...788...81M} {788, 81}

\bibitem[\protect\citeauthoryear{{Malo}, {Doyon}, {Feiden}, {Albert},
  {Lafreni{\`e}re}, {Artigau}, {Gagn{\'e}}  \& {Riedel}}{{Malo}
  et~al.}{2014b}]{2014b_Malo}
{Malo} L.,  {Doyon} R.,  {Feiden} G.~A.,  {Albert} L.,  {Lafreni{\`e}re} D.,
  {Artigau} {\'E}.,  {Gagn{\'e}} J.,   {Riedel} A.,  2014b, \mn@doi [\apj]
  {10.1088/0004-637X/792/1/37}, \href
  {http://adsabs.harvard.edu/abs/2014ApJ...792...37M} {792, 37}

\bibitem[\protect\citeauthoryear{{Mamajek}, {Meyer}, {Hinz}, {Hoffmann},
  {Cohen}  \& {Hora}}{{Mamajek} et~al.}{2004}]{2004a_Mamajek}
{Mamajek} E.~E.,  {Meyer} M.~R.,  {Hinz} P.~M.,  {Hoffmann} W.~F.,  {Cohen} M.,
    {Hora} J.~L.,  2004, \mn@doi [\apj] {10.1086/422550}, \href
  {https://ui.adsabs.harvard.edu/abs/2004ApJ...612..496M} {612, 496}

\bibitem[\protect\citeauthoryear{{Manara} et~al.,}{{Manara}
  et~al.}{2020}]{2020a_Manara}
{Manara} C.~F.,  et~al., 2020, \mn@doi [\aap] {10.1051/0004-6361/202037949},
  \href {https://ui.adsabs.harvard.edu/abs/2020A&A...639A..58M} {639, A58}

\bibitem[\protect\citeauthoryear{{Mart{\'{\i}}n}, {Lodieu}, {Pavlenko}  \&
  {B{\'e}jar}}{{Mart{\'{\i}}n} et~al.}{2018}]{2018a_Martin}
{Mart{\'{\i}}n} E.~L.,  {Lodieu} N.,  {Pavlenko} Y.,   {B{\'e}jar} V.~J.~S.,
  2018, \mn@doi [\apj] {10.3847/1538-4357/aaaeb8}, \href
  {http://adsabs.harvard.edu/abs/2018ApJ...856...40M} {856, 40}

\bibitem[\protect\citeauthoryear{{McMahon} et~al.,}{{McMahon}
  et~al.}{2019}]{2019a_McMahon}
{McMahon} et~al., 2019, VizieR Online Data Catalog, \href
  {https://ui.adsabs.harvard.edu/abs/2019yCat.2359....0M} {p. II/359}

\bibitem[\protect\citeauthoryear{{Monroe} \& {Pilachowski}}{{Monroe} \&
  {Pilachowski}}{2010}]{2010a_Monroe}
{Monroe} T.~R.,  {Pilachowski} C.~A.,  2010, \mn@doi [\aj]
  {10.1088/0004-6256/140/6/2109}, \href
  {https://ui.adsabs.harvard.edu/abs/2010AJ....140.2109M} {140, 2109}

\bibitem[\protect\citeauthoryear{{Morales} et~al.,}{{Morales}
  et~al.}{2009}]{2009a_Morales}
{Morales} J.~C.,  et~al., 2009, \mn@doi [\apj] {10.1088/0004-637X/691/2/1400},
  \href {https://ui.adsabs.harvard.edu/abs/2009ApJ...691.1400M} {691, 1400}

\bibitem[\protect\citeauthoryear{{Morin}, {Donati}, {Petit}, {Delfosse},
  {Forveille}  \& {Jardine}}{{Morin} et~al.}{2010}]{2010a_Morin}
{Morin} J.,  {Donati} J.~F.,  {Petit} P.,  {Delfosse} X.,  {Forveille} T.,
  {Jardine} M.~M.,  2010, \mn@doi [\mnras] {10.1111/j.1365-2966.2010.17101.x},
  \href {https://ui.adsabs.harvard.edu/abs/2010MNRAS.407.2269M} {407, 2269}

\bibitem[\protect\citeauthoryear{{Muzerolle}, {Calvet}, {Brice{\~n}o},
  {Hartmann}  \& {Hillenbrand}}{{Muzerolle} et~al.}{2000}]{2000a_Muzerolle}
{Muzerolle} J.,  {Calvet} N.,  {Brice{\~n}o} C.,  {Hartmann} L.,
  {Hillenbrand} L.,  2000, \mn@doi [\apjl] {10.1086/312691}, \href
  {https://ui.adsabs.harvard.edu/abs/2000ApJ...535L..47M} {535, L47}

\bibitem[\protect\citeauthoryear{{Naylor}}{{Naylor}}{2009}]{2009b_Naylor}
{Naylor} T.,  2009, \mn@doi [\mnras] {10.1111/j.1365-2966.2009.15295.x}, \href
  {https://ui.adsabs.harvard.edu/abs/2009MNRAS.399..432N} {399, 432}

\bibitem[\protect\citeauthoryear{{Palla}, {Randich}, {Pavlenko}, {Flaccomio}
  \& {Pallavicini}}{{Palla} et~al.}{2007}]{2007a_Palla}
{Palla} F.,  {Randich} S.,  {Pavlenko} Y.~V.,  {Flaccomio} E.,   {Pallavicini}
  R.,  2007, \mn@doi [\apjl] {10.1086/516733}, \href
  {http://adsabs.harvard.edu/abs/2007ApJ...659L..41P} {659, L41}

\bibitem[\protect\citeauthoryear{{Pasquini} et~al.,}{{Pasquini}
  et~al.}{2002}]{2002a_Pasquini}
{Pasquini} L.,  et~al., 2002, The Messenger, \href
  {https://ui.adsabs.harvard.edu/abs/2002Msngr.110....1P} {110, 1}

\bibitem[\protect\citeauthoryear{{Rajpurohit}, {Reyl{\'e}}, {Allard}, {Scholz},
  {Homeier}, {Schultheis}  \& {Bayo}}{{Rajpurohit}
  et~al.}{2014}]{2014a_Rajpurohit}
{Rajpurohit} A.~S.,  {Reyl{\'e}} C.,  {Allard} F.,  {Scholz} R.-D.,  {Homeier}
  D.,  {Schultheis} M.,   {Bayo} A.,  2014, \mn@doi [\aap]
  {10.1051/0004-6361/201322881}, \href
  {http://adsabs.harvard.edu/abs/2014A\%26A...564A..90R} {564, A90}

\bibitem[\protect\citeauthoryear{{Randich}, {Gilmore}  \& {Gaia-ESO
  Consortium}}{{Randich} et~al.}{2013}]{2013a_Randich}
{Randich} S.,  {Gilmore} G.,   {Gaia-ESO Consortium} 2013, The Messenger, \href
  {https://ui.adsabs.harvard.edu/abs/2013Msngr.154...47R} {154, 47}

\bibitem[\protect\citeauthoryear{{Randich} et~al.,}{{Randich}
  et~al.}{2018}]{2018a_Randich}
{Randich} S.,  et~al., 2018, \mn@doi [\aap] {10.1051/0004-6361/201731738},
  \href {https://ui.adsabs.harvard.edu/abs/2018A&A...612A..99R} {612, A99}

\bibitem[\protect\citeauthoryear{{Reiners} \& {Basri}}{{Reiners} \&
  {Basri}}{2009}]{2009a_Reiners}
{Reiners} A.,  {Basri} G.,  2009, \mn@doi [\apj]
  {10.1088/0004-637X/705/2/1416}, \href
  {http://adsabs.harvard.edu/abs/2009ApJ...705.1416R} {705, 1416}

\bibitem[\protect\citeauthoryear{{Rizzuto}, {Dupuy}, {Ireland}  \&
  {Kraus}}{{Rizzuto} et~al.}{2020}]{2020a_Rizzuto}
{Rizzuto} A.~C.,  {Dupuy} T.~J.,  {Ireland} M.~J.,   {Kraus} A.~L.,  2020,
  \mn@doi [\apj] {10.3847/1538-4357/ab5aed}, \href
  {https://ui.adsabs.harvard.edu/abs/2020ApJ...889..175R} {889, 175}

\bibitem[\protect\citeauthoryear{{Sacco} et~al.,}{{Sacco}
  et~al.}{2014}]{2014a_Sacco}
{Sacco} G.~G.,  et~al., 2014, \mn@doi [\aap] {10.1051/0004-6361/201423619},
  \href {https://ui.adsabs.harvard.edu/abs/2014A&A...565A.113S} {565, A113}

\bibitem[\protect\citeauthoryear{{Silverberg} et~al.,}{{Silverberg}
  et~al.}{2020}]{2020a_Silverberg}
{Silverberg} S.~M.,  et~al., 2020, \mn@doi [\apj] {10.3847/1538-4357/ab68e6},
  \href {https://ui.adsabs.harvard.edu/abs/2020ApJ...890..106S} {890, 106}

\bibitem[\protect\citeauthoryear{{Soderblom}}{{Soderblom}}{2010}]{2010a_Soderblom}
{Soderblom} D.~R.,  2010, \mn@doi [\araa]
  {10.1146/annurev-astro-081309-130806}, \href
  {http://adsabs.harvard.edu/abs/2010ARA\%26A..48..581S} {48, 581}

\bibitem[\protect\citeauthoryear{{Soderblom}, {Hillenbrand}, {Jeffries},
  {Mamajek}  \& {Naylor}}{{Soderblom} et~al.}{2013}]{2013a_Soderblom}
{Soderblom} D.~R.,  {Hillenbrand} L.~A.,  {Jeffries} R.~D.,  {Mamajek} E.~E.,
  {Naylor} T.,  2013, preprint, \href
  {http://adsabs.harvard.edu/abs/2013arXiv1311.7024S} {} (\mn@eprint {arXiv}
  {1311.7024})

\bibitem[\protect\citeauthoryear{{Somers} \& {Pinsonneault}}{{Somers} \&
  {Pinsonneault}}{2015}]{2015a_Somers}
{Somers} G.,  {Pinsonneault} M.~H.,  2015, \mn@doi [\mnras]
  {10.1093/mnras/stv630}, \href
  {http://adsabs.harvard.edu/abs/2015MNRAS.449.4131S} {449, 4131}

\bibitem[\protect\citeauthoryear{{Somers}, {Cao}  \& {Pinsonneault}}{{Somers}
  et~al.}{2020}]{2020a_Somers}
{Somers} G.,  {Cao} L.,   {Pinsonneault} M.~H.,  2020, \mn@doi [\apj]
  {10.3847/1538-4357/ab722e}, \href
  {https://ui.adsabs.harvard.edu/abs/2020ApJ...891...29S} {891, 29}

\bibitem[\protect\citeauthoryear{{Spina} et~al.,}{{Spina}
  et~al.}{2017}]{2017a_Spina}
{Spina} L.,  et~al., 2017, \mn@doi [\aap] {10.1051/0004-6361/201630078}, \href
  {https://ui.adsabs.harvard.edu/abs/2017A&A...601A..70S} {601, A70}

\bibitem[\protect\citeauthoryear{{Stauffer}, {Schultz}  \&
  {Kirkpatrick}}{{Stauffer} et~al.}{1998}]{1998a_Stauffer}
{Stauffer} J.~R.,  {Schultz} G.,   {Kirkpatrick} J.~D.,  1998, \mn@doi [\apjl]
  {10.1086/311379}, \href {http://adsabs.harvard.edu/abs/1998ApJ...499L.199S}
  {499, L199}

\bibitem[\protect\citeauthoryear{{Stauffer} et~al.,}{{Stauffer}
  et~al.}{1999}]{1999a_Stauffer}
{Stauffer} J.~R.,  et~al., 1999, \mn@doi [\apj] {10.1086/308069}, \href
  {http://adsabs.harvard.edu/abs/1999ApJ...527..219S} {527, 219}

\bibitem[\protect\citeauthoryear{{Tognelli}, {Prada Moroni}  \&
  {Degl'Innocenti}}{{Tognelli} et~al.}{2015}]{2015a_Tognelli}
{Tognelli} E.,  {Prada Moroni} P.~G.,   {Degl'Innocenti} S.,  2015, \mn@doi
  [\mnras] {10.1093/mnras/stv577}, \href
  {https://ui.adsabs.harvard.edu/abs/2015MNRAS.449.3741T} {449, 3741}

\bibitem[\protect\citeauthoryear{{Torres}}{{Torres}}{2013}]{2013a_Torres}
{Torres} G.,  2013, \mn@doi [Astronomische Nachrichten]
  {10.1002/asna.201211743}, \href
  {https://ui.adsabs.harvard.edu/abs/2013AN....334....4T} {334, 4}

\makeatother
\end{thebibliography}

\appendix

\section{The metallicity of NGC~2232}\label{S_Metallicity}

There are 14 NGC~2232 members with [Fe/H] measured in GESiDR5 that have corresponding uncertainties and are therefore deemed reliable measurements, all of which were obtained using the GIRAFFE spectrograph. These are listed in Table~\ref{T_Metallicities}, where the $T_{\rm eff}$ indicate the majority of these stars are of F-K-type. Both the weighted mean [Fe/H] for these targets and the metallicity range across individual stars suggests the metallicity of NGC~2232 ([Fe/H]\ $=0.00 \pm 0.014$) is entirely consistent with a solar value, and similar to the metallicity of most nearby, young open clusters observed in GES (\citealt{2017a_Spina}). No evidence of a super-Solar [Fe/H] is found, as reported by \cite{2010a_Monroe}.

{\centering
\begin{table}
\begin{center}
\begin{tabular}{lrrr}
\hline
\hline
GES Name & SNR & [Fe/H] & $T_{\rm eff}$ \\
(GES J-) &     & dex    & (K) \\
\hline
06283370$-$0446583 & $121$ & $-0.094 \pm 0.034$ & $5797 \pm	193$ \\

06250357$-$0456194 & $181$ & $-0.083 \pm 0.076$ & $6316 \pm	121$ \\
06262350$-$0416106 & $48$  & $-0.035 \pm 0.073$ & $4243 \pm	140$ \\
06293045$-$0441236 & $137$ & $-0.034 \pm 0.008$ & $4934 \pm	161$ \\
06281077$-$0432425 & $91$  & $-0.033 \pm 0.037$ & $4503 \pm	4$   \\
06284806$-$0442437 & $88$  & $-0.006 \pm 0.020$ & $4988 \pm	99 $ \\
06274602$-$0446224 & $196$ & $+0.002 \pm 0.081$ & $5978 \pm	52$  \\
06281883$-$0450578 & $52$  & $+0.004 \pm 0.163$ & $4046 \pm	116$ \\
06274774$-$0440331 & $134$ & $+0.013 \pm 0.054$ & $5785 \pm	123$ \\
06254936$-$0449006 & $76$  & $+0.020 \pm 0.092$ & $4760 \pm	17$  \\
06285630$-$0449096 & $123$ & $+0.035 \pm 0.021$ & $5181 \pm	198$ \\
06262866$-$0437367 & $153$ & $+0.068 \pm 0.002$ & $5970 \pm	41$  \\
06262777$-$0433218 & $177$ & $+0.071 \pm 0.105$ & $6054 \pm	245$ \\
06251933$-$0442048 & $135$ & $+0.077 \pm 0.023$ & $5468 \pm	122$ \\
\hline
\multicolumn{4}{c}{{\bf Final [Fe/H] = $0.000 \pm 0.054 \pm 0.014$\,dex}} \\
\hline
\end{tabular}
\caption{Metallicity values for the 14 NGC~2232 targets observed with the GIRAFFE spectrograph. The values in this table are those reported in the GESiDR5 catalogue. The final [Fe/H] value is the mean value, and the two error bars are the standard deviation and standard error in the mean, respectively.}
\label{T_Metallicities}
\end{center}
\end{table}}

\section{Using empirical continuum spectra to estimate EW(Li) for K and M-dwarf stars}\label{S_EWLiMethod}

This section describes the method used to estimate the EW(Li) of K and M-dwarfs in NGC~2232 using the GESiDR5 spectra observed with the GIRAFFE spectrograph and 665nm (HR15n) filter. The method proceeds in four steps.

\subsection{Sample selection}
\cite{2020a_Jackson}, hereafter J20, used GESiDR5 data to estimate  membership probabilities for 10817 HR15n targets in 32 open clusters.  Of these, 4390 were identified as likely cluster members with membership probability $P_{\rm 3D}>0.9$ and 5726 as likely field stars with $P_{\rm 3D}<0.1$ (see Table 3 of J20). Spectra of these targets were analysed to determine their spectral indices using the procedure described in \cite{2014a_Damiani}. In particular, the temperature-sensitive $\tau$ index was calculated to estimate $(G-K_{\rm s})_0$ for dwarf field stars of unknown reddening. 

\begin{figure}
  \centering
	\includegraphics[width=0.48\textwidth]{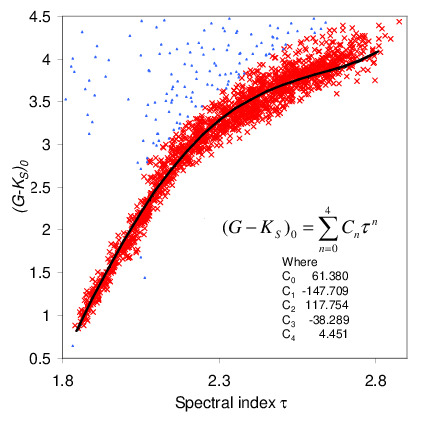}
	\caption{$(G-K_{\rm s})_0$ versus temperature index $\tau$ for $P_{\rm 3D}> 0.9$ cluster members with Gaia DR2 parallax $>1$'' (see table 3 of J20). The black curve shows a polynomial fit to the measured data using coefficients shown on the plot. Red crosses are data used to fit the curve, blue triangles are outliers excluded from the fit.}
	\label{tau2GK}
\end{figure}

A subset of the data corresponding to likely dwarf cluster members with Gaia DR2 parallaxes $>1$~mas was used to determine a fourth order polynomial relation of  $(G-K_{\rm s})_0$ as a function $\tau$ over the range $1.85<\tau<2.8$ using the cluster reddenings shown in table 1 of J20. The polynomial fit, shown in Figure~\ref{tau2GK}, was then used to estimate $(G-K_{\rm s})_0$ values for the sample of likely field stars.

\subsection{Empirical continuum spectra}
A set of dwarf field stars were selected with good $G$ and $K_{\rm s}$ photometry, a parallax $>1$\,mas and low levels of EW(Li) reported in GESiDR5 (EW(Li) $< 50$\,m\AA~for targets with $(G-K_{\rm s})_0<3.6$ and EW(Li) $<200\,$m\AA~for cooler stars). Targets were binned by $(G-K_{\rm s})_0$ colour in $\pm 0.1$\,mag wide bins in steps of 0.1\,mag between $1.6 <(G-K_{\rm s})_0<4.0$.

The spectra were offset to their rest wavelength scale using the GESiDR5 RV and normalised to their median value over the wavelength range 6675--6730\,\AA. Empirical continuum spectra were calculated as the median of the normalised spectra in each bin at each point over the  wavelength range. The uncertainty in the continuum spectrum was estimated as 1.3 times the median absolute deviation (MAD) of the sample spectra relative to the median continuum spectrum.

\begin{figure*}
	\includegraphics[width=0.9\linewidth]{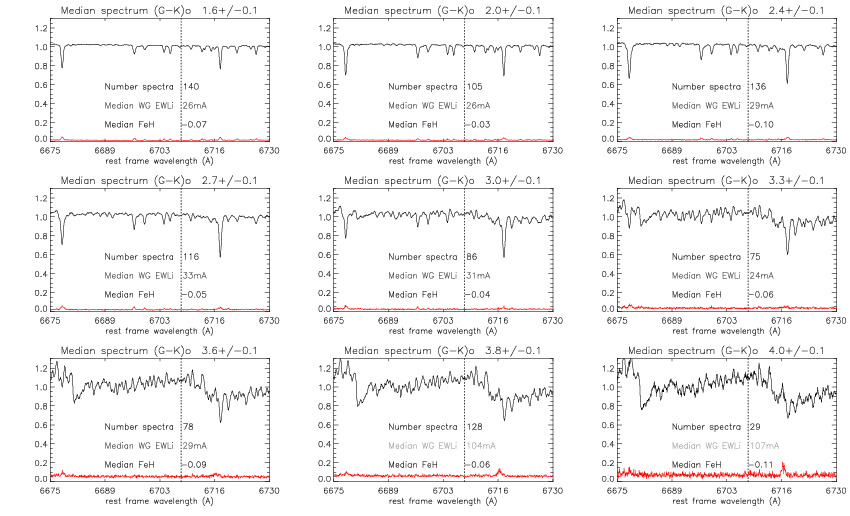}
	\caption{Examples of median continuum spectra for stars with low levels of Li as a function of $(G-K_{\rm s})_0$. Text on each plot shows the number of spectra averaged in each bin and median values of EW(Li) and [Fe/H]. The lower red curve shows the uncertainty in the continuum waveform, calculated as 1.3 times the median absolute deviation of individual spectra from the median spectrum. The dotted line indicates the position of the Li\,{\sc i}~6708\,\AA\ absorption line.}
	\label{continuum}	
\end{figure*}

Examples of continuum spectra and their associated uncertainties are shown in Figure~\ref{continuum} for $1.6< (G-K_{\rm s})_0<4.0$. The spectra of late M-dwarfs change considerably due to developing molecular absorption. Uncertainties in the continuum spectra increase from $\sim 1$ per cent for the warmest stars to $\sim 7$ per cent at $(G-K_{\rm s})_0=4$. These uncertainties were used to estimate the additional uncertainties in EW(Li) due to potential mismatch between the median continuum spectra and an ``ideal'' continuum spectrum for a particular target. 

It is recognised that some of the targets used to define these median spectra may not be fully Li-depleted. However, in M-dwarfs the depletion of Li takes place very rapidly and there should be few field M-dwarfs with low, but non-zero EW(Li). This is supported by the absence of any sign of a lithium feature in any of the median spectra.

\subsection{FWHM of target spectra in NGC~2232}  
An optimal extraction algorithm \citep{1986a_Horne} was used to measure EW(Li) where the FWHM of the extraction profile is a function of the spectral resolution and target $v\sin i$. The GES pipeline used to estimate rotational velocities for GESiDR5 data assumes a fixed spectral resolving power of $R = 15\,000$ for filter HR15n. However, over the period when NGC~2232 was observed the resolution measured from the line width of daily arc-lamp spectra was unusually low, varying between $R=12075$ and $R=12534$ (J20). For this reason the GESiDR5 rotational velocities ({\tt VROT}) were corrected to give a reduced value, $v\sin i$, as detailed in appendix~A of J20. The FWHM of the extraction profile was calculated as:
\begin{eqnarray}
	{\rm FWHM} & = & \frac{\lambda_{\rm Li}}{ R}\left[1+\frac{(v\sin i)^2}{C^2}\right];\\\nonumber
	{\rm where\,\,\,} C & = &\left[\frac{1-u/3}{1-7u/15}\right]^\frac{1}{2}\frac{c}{R\sqrt{2\ln2}}\, , \nonumber
\end{eqnarray}
where $\lambda_{\rm Li}$ is the wavelength of the Li line at 6708\,\AA\ and $R$ is the resolving power on the day the target was observed. The term $(v\sin i)^2/C^2$ represents the effect of rotational broadening, taken from \cite{1984a_Gray}, for a star with limb darkening coefficient $u$ \citep[see][appendix A]{2015a_Jackson}.

\subsection{Measurement of EW(Li)}
EW(Li) was measured by comparing target spectra (corrected to a rest wavelength scale) to the median continuum spectrum matched to the target $(G-K_{\rm s})_0$ colour and scaled to match the continuum spectrum either side of the Li line over wavelength ranges, 6685--6705\AA~and 6711--6717\,\AA. A weighted profile $P(\lambda$) was used to estimate EW(Li) from the difference between the target ($S_{T}(\lambda)$) and template ($S_{C}(\lambda)$) spectra;
\begin{equation}
	{\rm EW(Li)} = \int [S_{C}(\lambda)-S_{T}(\lambda)]P(\lambda)\, d\lambda \ / \int
	P(\lambda)^2\, d\lambda\, ,
\end{equation}
where $P(\lambda)$ is a Gaussian profile with the predicted FWHM of the target spectrum. The uncertainty in EW(Li) was estimated as the RMS value of the EWs measured using the same procedure with $P(\lambda)$ centred at 15 offset wavelengths to the left of the Li line and 5 to the right. The systematic uncertainty in EW(Li) was estimated as a function of the Gaussian profile and the uncertainty spectrum at the appropriate $(G-K_{\rm s})_0$ colour. Results are shown in Table~A1 and Figure~\ref{fig:ewli_gks_rjj}.

\section{Data for all NGC~2232 targets}

Table~\ref{T_Data} presents the relevant data for all 80 stars that constitute our target sample.

\clearpage
\onecolumn
\begin{center}
\begin{longtable}{p{2.9cm}p{0.65cm}p{2.25cm}p{2.25cm}p{2.25cm}p{1.6cm}p{2.0cm}p{0.7cm}}
\caption{Gaia DR2 source identifiers (column 1), membership probabilities ($P_{\rm 3D}$, column 2), photometry (columns 3-5, before applying corrections for extinction and reddening), EW(Li) values reported in the GESiDR5 catalogue (column 6) and our analysis in $\S$\ref{S_EWLi_us} (column 7) and the signal-to-noise ratio of the GESiDR5 spectra.} \\
\hline
\hline
Gaia DR2            & $P_{\rm 3D}$ & $G$           & $G-J$    & $G-K_{\rm s}$ & EW(Li)DR5 & EW(Li)RJJ & SNR \\
\hline
3104419133401411840 & $0.999$ & $11.6590 \pm 0.0003$ & $0.8930 \pm 0.0240$ & $1.1990 \pm 0.0190$ & $118 \pm 3$  & $+103.3 \pm 11.2$ & $181.5$ \\
3104457096612037504 & $0.998$ & $12.1306 \pm 0.0008$ & $1.0046 \pm 0.0300$ & $1.3266 \pm 0.0320$ & $170 \pm 9$  & $+148.8 \pm 5.7$ & $177.2$ \\
3104435346901124992 & $1.000$ & $12.0718 \pm 0.0025$ & $1.0608 \pm 0.0230$ & $1.4228 \pm 0.0230$ & $184 \pm 6$  & $+175.3 \pm 5.1$ & $195.9$ \\
3104454347832973696 & $0.998$ & $12.3488 \pm 0.0006$ & $1.0748 \pm 0.0260$ & $1.4268 \pm 0.0230$ & $157 \pm 3$  & $+139.0 \pm 3.7$ & $153.3$ \\
3104532761057366016 & $1.000$ & $12.1121 \pm 0.0010$ & $0.9961 \pm 0.0290$ & $1.4551 \pm 0.0230$ & $197 \pm 4$  & $+177.9 \pm 3.7$ & $133.8$ \\
3104339964265249536 & $1.000$ & $12.1526 \pm 0.0005$ & $1.1186 \pm 0.0260$ & $1.5536 \pm 0.0230$ & $155$        & $+145.0 \pm 3.8$ & $121.5$ \\
3104494724825684608 & $0.999$ & $12.3779 \pm 0.0025$ & $1.3749 \pm 0.0270$ & $1.8639 \pm 0.0260$ & $257$        & $+233.7 \pm 6.6$ & $134.6$ \\
3104336562651152128 & $0.999$ & $12.7301 \pm 0.0019$ & $1.3470 \pm 0.0260$ & $1.8920 \pm 0.0240$ & $292 \pm 4$  & $+272.5 \pm 7.4$ & $122.6$ \\
3104345255664954368 & $0.982$ & $12.2941 \pm 0.0035$ & $1.3761 \pm 0.0230$ & $1.9231 \pm 0.0230$ & $289 \pm 4$  & $+276.1 \pm 10.2$ & $136.6$ \\
3104341858349761152 & $0.999$ & $12.8421 \pm 0.0024$ & $1.4621 \pm 0.0240$ & $2.0541 \pm 0.0230$ & $316 \pm 5$  & $+273.2 \pm 5.7$ & $88.2$ \\
3104444868845867264 & $0.996$ & $13.5797 \pm 0.0047$ & $1.6617 \pm 0.0260$ & $2.3297 \pm 0.0300$ & $386 \pm 6$  & $+354.4 \pm 7.2$ & $76.3$ \\
3104535784714244480 & $1.000$ & $13.9327 \pm 0.0015$ & $1.7447 \pm 0.0340$ & $2.4927 \pm 0.0300$ & $363 \pm 4$  & $+322.7 \pm 3.8$ & $91.0$ \\
3104605191385258112 & $1.000$ & $14.4590 \pm 0.0024$ & $2.0940 \pm 0.0260$ & $2.9220 \pm 0.0240$ & $357 \pm 7$  & $+315.4 \pm 6.3$ & $48.5$ \\
3104245543704247168 & $1.000$ & $14.8786 \pm 0.0028$ & $1.7788 \pm 0.0011$ & $3.1886 \pm 0.0240$ & $478 \pm 8$  & $+424.6 \pm 10.8$ & $51.6$ \\
3104342579904273152 & $0.987$ & $14.5603 \pm 0.0018$ & $2.4343 \pm 0.0260$ & $3.3093 \pm 0.0230$ & $354 \pm 13$ & $+289.4 \pm 12.8$ & $32.2$ \\
3104535990872706304 & $1.000$ & $15.3370 \pm 0.0017$ & $2.4742 \pm 0.0010$ & $3.3220 \pm 0.0190$ & $122 \pm 6$  & $+36.6 \pm 11.7$ & $29.0$ \\
3104245857239589632 & $1.000$ & $15.3893 \pm 0.0035$ & $2.1241 \pm 0.0012$ & $3.3253 \pm 0.0230$ & $89 \pm 7$   & $+2.0 \pm 5.1$ & $83.2$ \\
3104431812145190272 & $1.000$ & $15.3449 \pm 0.0023$ & $2.1768 \pm 0.0012$ & $3.3389 \pm 0.0240$ & $136 \pm 22$ & $+85.5 \pm 9.4$ & $37.2$ \\
3104235892915628416 & $1.000$ & $15.5812 \pm 0.0011$ & $2.3148 \pm 0.0012$ & $3.3492 \pm 0.0230$ & $120 \pm 11$ & $+25.5 \pm 7.1$ & $49.1$ \\
3104529973619388672 & $1.000$ & $15.5582 \pm 0.0013$ & $2.5955 \pm 0.0011$ & $3.3752 \pm 0.0230$ & $371 \pm 23$ & $+306.7 \pm 9.6$ & $46.4$ \\
3104238950932224512 & $1.000$ & $15.4832 \pm 0.0021$ & $2.3162 \pm 0.0012$ & $3.3852 \pm 0.0230$ & $161 \pm 13$ & $+97.8 \pm 8.2$ & $44.8$ \\
3104246681873275008 & $1.000$ & $15.7425 \pm 0.0014$ & $2.3493 \pm 0.0013$ & $3.3855 \pm 0.0260$ & $75$         & $-4.9 \pm 12.5$ & $37.5$ \\
3104323029213231488 & $1.000$ & $15.4756 \pm 0.0018$ & $2.4003 \pm 0.0011$ & $3.3956 \pm 0.0240$ & $128 \pm 12$ & $+41.0 \pm 8.2$ & $47.0$ \\
3104601995929583360 & $1.000$ & $15.6269 \pm 0.0015$ & $2.6373 \pm 0.0012$ & $3.4679 \pm 0.0270$ & $133 \pm 10$ & $+4.8 \pm 15.3$ & $20.3$ \\
3104431949584129280 & $1.000$ & $15.7826 \pm 0.0023$ & $2.4707 \pm 0.0013$ & $3.4706 \pm 0.0240$ & $118 \pm 14$ & $-4.5 \pm 9.6$ & $45.3$ \\
3104446037076422912 & $1.000$ & $14.8382 \pm 0.0011$ & $2.5962 \pm 0.0260$ & $3.5042 \pm 0.0210$ & $66$         & $+261.9 \pm 27.4$ & $33.6$ \\
3104534960080775040 & $1.000$ & $15.1646 \pm 0.0017$ & $2.6096 \pm 0.0240$ & $3.5326 \pm 0.0230$ & $95 \pm 10$  & $-4.8 \pm 7.9$ & $55.6$ \\
3104528328651278208 & $0.987$ & $15.7663 \pm 0.0009$ & $2.4520 \pm 0.0013$ & $3.5733 \pm 0.0260$ & $94 \pm 10$  & $-16.6 \pm 17.7$ & $28.7$ \\
3104558732721761920 & $0.999$ & $16.2700 \pm 0.0025$ & $2.7422 \pm 0.0016$ & $3.5910 \pm 0.0260$ & $121 \pm 6$  & $-22.6 \pm 14.0$ & $29.9$ \\
3104214521158310144 & $0.999$ & $16.1659 \pm 0.0026$ & $2.7343 \pm 0.0014$ & $3.6019 \pm 0.0270$ & $1$          & $-31.3 \pm 16.7$ & $14.8$ \\
3104432671138660480 & $1.000$ & $16.3635 \pm 0.0014$ & $2.7411 \pm 0.0015$ & $3.6175 \pm 0.0300$ & $132$        & $-37.5 \pm 18.3$ & $28.3$ \\
3104225997310909696 & $0.999$ & $16.3812 \pm 0.0016$ & $2.9444 \pm 0.0014$ & $3.6601 \pm 0.0290$ & $87 \pm 13$  & $-14.5 \pm 21.8$ & $15.0$ \\
3104528466090065024 & $1.000$ & $16.2735 \pm 0.0021$ & $2.8704 \pm 0.0013$ & $3.6674 \pm 0.0230$ & $52 \pm 11$  & $-16.4 \pm 18.2$ & $25.5$ \\
3104455417284845440 & $1.000$ & $15.6339 \pm 0.0022$ & $2.7469 \pm 0.0010$ & $3.6759 \pm 0.0240$ & $165 \pm 13$ & $+29.2 \pm 16.2$ & $20.8$ \\
3104173525693960704 & $0.999$ & $16.2391 \pm 0.0020$ & $2.8244 \pm 0.0013$ & $3.6881 \pm 0.0260$ & $134 \pm 16$ & $-10.0 \pm 27.5$ & $11.9$ \\
3104605019586592256 & $1.000$ & $16.2785 \pm 0.0012$ & $2.8393 \pm 0.0015$ & $3.6885 \pm 0.0240$ & $153 \pm 19$ & $-25.5 \pm 18.7$ & $10.9$ \\
3104454901887364736 & $0.999$ & $16.2260 \pm 0.0024$ & $2.8556 \pm 0.0013$ & $3.6940 \pm 0.0350$ & $155 \pm 22$ & $+4.4 \pm 26.5$ & $13.2$ \\
3104506922532654848 & $0.996$ & $15.9396 \pm 0.0024$ & $3.0363 \pm 0.0010$ & $3.7396 \pm 0.0260$ & $10$         & $-1.3 \pm 48.9$ & $17.5$ \\
3104474108982725888 & $1.000$ & $16.6415 \pm 0.0026$ & $2.9218 \pm 0.0016$ & $3.7435 \pm 0.0300$ & $125 \pm 10$ & $+27.3 \pm 17.1$ & $19.9$ \\
3104434286046639744 & $0.998$ & $16.4142 \pm 0.0033$ & $2.9474 \pm 0.0014$ & $3.7642 \pm 0.0230$ & $100 \pm 20$ & $+2.8 \pm 12.4$ & $30.7$ \\
3104528289992577152 & $0.998$ & $17.0776 \pm 0.0019$ & $2.9760 \pm 0.0020$ & $3.7940 \pm 0.0033$ & $84 \pm 21$  & $+7.8 \pm 19.3$ & $19.8$ \\
3104226306548544128 & $0.997$ & $15.9935 \pm 0.0023$ & $2.8783 \pm 0.0011$ & $3.7955 \pm 0.0260$ & $77$         & $-45.4 \pm 23.8$ & $30.1$ \\
3104425588734811648 & $0.998$ & $16.9580 \pm 0.0018$ & $2.9737 \pm 0.0019$ & $3.7990 \pm 0.0031$ & $99 \pm 21$  & $+14.8 \pm 33.6$ & $10.5$ \\
3104243589496954624 & $1.000$ & $16.4457 \pm 0.0023$ & $2.8700 \pm 0.0015$ & $3.8087 \pm 0.0330$ & $112 \pm 17$ & $-3.0 \pm 10.8$ & $32.1$ \\
3104440814396743552 & $0.999$ & $16.9054 \pm 0.0026$ & $2.9961 \pm 0.0018$ & $3.8128 \pm 0.0029$ & $65 \pm 13$  & $-38.8 \pm 19.6$ & $16.6$ \\
3104454317773799808 & $0.999$ & $17.1865 \pm 0.0017$ & $3.0133 \pm 0.0021$ & $3.8266 \pm 0.0035$ & $79$         & $+3.3 \pm 36.3$ & $6.0$ \\
3104506437197663232 & $1.000$ & $16.9640 \pm 0.0017$ & $3.0283 \pm 0.0018$ & $3.8482 \pm 0.0030$ & $176$        & $+28.5 \pm 39.3$ & $9.0$ \\
3104246578794084608 & $0.999$ & $17.2219 \pm 0.0016$ & $3.0701 \pm 0.0021$ & $3.8646 \pm 0.0035$ & $76$         & $-68.3 \pm 42.5$ & $8.2$ \\
3104582376519289600 & $0.999$ & $17.0763 \pm 0.0026$ & $3.0403 \pm 0.0022$ & $3.8652 \pm 0.0035$ & $107 \pm 17$ & $-42.2 \pm 29.8$ & $12.2$ \\
3104240840717772288 & $0.999$ & $17.0603 \pm 0.0015$ & $3.0392 \pm 0.0019$ & $3.8654 \pm 0.0031$ & $66$         & $-13.1 \pm 27.7$ & $12.5$ \\
3104536952945322880 & $1.000$ & $16.4957 \pm 0.0026$ & $3.0116 \pm 0.0014$ & $3.8657 \pm 0.0260$ & $103 \pm 9$  & $+14.4 \pm 14.5$ & $21.8$ \\
3104530424595112576 & $1.000$ & $16.4198 \pm 0.0035$ & $3.0508 \pm 0.0013$ & $3.8858 \pm 0.0300$ & $80 \pm 9$   & $+9.1 \pm 20.5$ & $22.8$ \\
3104440676957804672 & $0.998$ & $16.6423 \pm 0.0029$ & $3.0207 \pm 0.0015$ & $3.8893 \pm 0.0300$ & $67 \pm 11$  & $-3.1 \pm 20.7$ & $19.2$ \\
3104438787171978624 & $1.000$ & $16.7697 \pm 0.0037$ & $3.0383 \pm 0.0016$ & $3.8907 \pm 0.0320$ & $134$        & $-2.7 \pm 16.0$ & $20.4$ \\
3104331722226892800 & $0.998$ & $17.4389 \pm 0.0017$ & $3.1217 \pm 0.0023$ & $3.9154 \pm 0.0039$ & $230$        & $-4.3 \pm 51.2$ & $6.4$ \\
3104425970989680512 & $0.994$ & $17.4495 \pm 0.0016$ & $3.1134 \pm 0.0023$ & $3.9295 \pm 0.0040$ & $98$         & $-50.7 \pm 46.8$ & $10.4$ \\
3104505857380683776 & $0.999$ & $16.7492 \pm 0.0012$ & $3.1061 \pm 0.0015$ & $3.9362 \pm 0.0270$ & $232$        & $+33.8 \pm 45.5$ & $9.5$ \\
3104504341253607552 & $0.998$ & $17.2799 \pm 0.0014$ & $3.1121 \pm 0.0021$ & $3.9366 \pm 0.0035$ & $117 \pm 33$ & $+40.6 \pm 47.5$ & $7.2$ \\
3104331477410631552 & $0.999$ & $17.8006 \pm 0.0024$ & $3.1594 \pm 0.0029$ & $3.9812 \pm 0.0049$ & $189$        & $+46.3 \pm 45.5$ & $7.2$ \\
3104341136795325312 & $0.977$ & $16.6180 \pm 0.0011$ & $3.1559 \pm 0.0014$ & $4.0120 \pm 0.0260$ & $142$        & $-57.9 \pm 32.3$ & $20.3$ \\
3104551414093862400 & $0.993$ & $18.0674 \pm 0.0027$ & $3.2488 \pm 0.0032$ & $4.0461 \pm 0.0058$ & $82$         & $+49.7 \pm 77.5$ & $3.3$ \\
3104423668887274496 & $0.997$ & $17.7533 \pm 0.0041$ & $3.2266 \pm 0.0026$ & $4.0543 \pm 0.0046$ & $283$        & $-308.6 \pm 175.7$ & $3.4$ \\
3104434625346741504 & $0.997$ & $18.1226 \pm 0.0032$ & $3.2395 \pm 0.0035$ & $4.1058 \pm 0.0058$ & $245$        & $-64.6 \pm 41.8$ & $6.6$ \\
3104214482502076672 & $0.999$ & $18.2245 \pm 0.0028$ & $3.2967 \pm 0.0034$ & $4.1080 \pm 0.0062$ & $301$        & $+101.6 \pm 69.9$ & $4.6$ \\
3104455619144457600 & $0.952$ & $18.6173 \pm 0.0036$ & $3.3926 \pm 0.0043$ & $4.1292 \pm 0.0087$ &              & $+550.9 \pm 235.3$ & $1.7$ \\
3104438061320152576 & $0.997$ & $18.2611 \pm 0.0025$ & $3.3256 \pm 0.0035$ & $4.1337 \pm 0.0063$ & $301$        & $+26.3 \pm 163.6$ & $2.3$ \\
3104329999947214336 & $0.999$ & $18.2442 \pm 0.0030$ & $3.3485 \pm 0.0034$ & $4.1557 \pm 0.0062$ & $16$         & $-99.8 \pm 61.7$ & $4.7$ \\
3104328045734809088 & $0.998$ & $18.4632 \pm 0.0035$ & $3.3488 \pm 0.0040$ & $4.1726 \pm 0.0072$ & $371$        & $+43.5 \pm 113.2$ & $3.6$ \\
3104212558354814464 & $0.988$ & $17.6424 \pm 0.0021$ & $3.3891 \pm 0.0022$ & $4.1906 \pm 0.0037$ & $340$        & $+158.1 \pm 47.2$ & $6.6$ \\
3104243756998337152 & $0.989$ & $18.3675 \pm 0.0034$ & $3.3856 \pm 0.0036$ & $4.2223 \pm 0.0065$ & $118$        & $+161.3 \pm 126.1$ & $2.8$ \\
3104434316109083136 & $0.999$ & $18.2833 \pm 0.0031$ & $3.4164 \pm 0.0033$ & $4.2395 \pm 0.0059$ & $319$        & $+11.2 \pm 107.9$ & $4.2$ \\
3104433869432346240 & $0.997$ & $17.6949 \pm 0.0019$ & $3.4324 \pm 0.0022$ & $4.2560 \pm 0.0037$ & $377$        & $+159.6 \pm 46.3$ & $7.2$ \\
3104243860077501056 & $0.993$ & $18.7621 \pm 0.0043$ & $3.4224 \pm 0.0046$ & $4.2576 \pm 0.0087$ &              & $+538.5 \pm 182.7$ & $1.5$ \\
3104431124950743936 & $0.988$ & $17.8219 \pm 0.0037$ & $3.4325 \pm 0.0024$ & $4.2612 \pm 0.0041$ & $239 \pm 30$ & $+515.2 \pm 97.0$ & $6.2$ \\
3104561893816700800 & $0.990$ & $18.9278 \pm 0.0046$ & $3.4628 \pm 0.0061$ & $4.2892 \pm 0.0108$ & $843$        & $+732.7 \pm 159.7$ & $2.5$ \\
3104426482088556800 & $0.995$ & $18.7511 \pm 0.0041$ & $3.5029 \pm 0.0044$ & $4.3109 \pm 0.0083$ & $342 \pm 41$ & $+328.1 \pm 100.3$ & $3.0$ \\
3104341132499390336 & $0.982$ & $19.1105 \pm 0.0055$ & $3.5576 \pm 0.0054$ & $4.3786 \pm 0.0106$ & $827$        & $+318.0 \pm 242.0$ & $1.3$ \\
3104216204782370048 & $0.994$ & $16.3720 \pm 0.0093$ & $3.4406 \pm 0.0010$ & $4.3980 \pm 0.0240$ & $134 \pm 14$ & $+63.4 \pm 25.9$ & $15.2$ \\
3104534715263592192 & $0.991$ & $18.5571 \pm 0.0032$ & $3.5934 \pm 0.0036$ & $4.4142 \pm 0.0065$ & $675$        & $+470.9 \pm 166.4$ & $1.9$ \\
3104340891981168128 & $0.989$ & $19.2257 \pm 0.0045$ & $3.6745 \pm 0.0054$ & $4.5418 \pm 0.0100$ & $756$        & $+511.2 \pm 240.1$ & $1.6$ \\
\hline
\label{T_Data}

\end{longtable}
\end{center}
\clearpage
\twocolumn

\label{lastpage}

\end{document}